\documentclass{article}

\usepackage{amssymb}
\usepackage{arxiv}
\usepackage{graphicx}
\usepackage[utf8]{inputenc} % allow utf-8 input
\usepackage[T1]{fontenc}    % use 8-bit T1 fonts
\usepackage{hyperref}       % hyperlinks
\usepackage{url}            % simple URL typesetting
\usepackage{booktabs}       % professional-quality tables
\usepackage{amsfonts} 
\usepackage{amssymb}% blackboard math symbols
\usepackage{nicefrac}       % compact symbols for 1/2, etc.
\usepackage{microtype}      % microtypography
\usepackage{lipsum}
\usepackage{bm}
\usepackage{xspace}

\usepackage{amsmath}
\usepackage{cases}
\usepackage{mathtools}
\usepackage[ruled,vlined]{algorithm2e}
\usepackage[capitalise]{cleveref} %must be loaded after amsmath to avoid issues
\usepackage{appendix}

\usepackage{empheq}
\newcommand*\widefbox[1]{\fbox{\hspace{2em}#1\hspace{2em}}}

\usepackage{setspace} %for double spacing
\newcommand{\bEd}{\ensuremath{\dot{\mathbf{E}}}\xspace}

\newcommand{\bA}{\ensuremath{\mathbf{A}}\xspace}
\newcommand{\bB}{\ensuremath{\mathbf{B}}\xspace}
\newcommand{\bD}{\ensuremath{\mathbf{D}}\xspace}
\newcommand{\bC}{\ensuremath{\mathbf{C}}\xspace}
\newcommand{\bE}{\ensuremath{\mathbf{E}}\xspace}
\newcommand{\bF}{\ensuremath{\mathbf{F}}\xspace}

\newcommand{\bI}{\ensuremath{\mathbf{I}}\xspace}
\newcommand{\bL}{\ensuremath{\mathbf{L}}\xspace}

\newcommand{\bQ}{\ensuremath{\mathbf{Q}}\xspace}
\newcommand{\bT}{\ensuremath{\mathbf{T}}\xspace}

\newcommand{\bW}{\ensuremath{\mathbf{W}}\xspace}
\newcommand{\bZ}{\ensuremath{\mathbf{Z}}\xspace}

\newcommand{\bd}{\ensuremath{\mathbf{d}}\xspace}
\newcommand{\bl}{\ensuremath{\mathbf{l}}\xspace}

\newcommand{\bn}{\ensuremath{\mathbf{n}}\xspace}
\newcommand{\bq}{\ensuremath{\mathbf{q}}\xspace}
\newcommand{\br}{\ensuremath{\mathbf{r}}\xspace}

\newcommand{\bt}{\ensuremath{\mathbf{t}}\xspace}
\newcommand{\bu}{\ensuremath{\mathbf{u}}\xspace}

\newcommand{\bx}{\ensuremath{\mathbf{x}}\xspace}

\newcommand{\sz}{_{[0]}}
\newcommand{\so}{_{[1]}}

\newcommand{\tDt}{{\tau+\Delta t}}
\newcommand{\Dt}{{\Delta t}}

\newcommand{\bLambda}{\ensuremath{\mathbf{\Lambda}}\xspace}

    {\begin{equation}
    \begin{gathered}
    }
    { 
    \end{gathered} 
    \end{equation}
    }

    {\begin{align}
    }
    { 
    \end{align}
    }
    
    {\begin{bmatrix}
    }
    { 
    \end{bmatrix}
    }

\let\originalleft\left %see https://tex.stackexchange.com/questions/2607/spacing-around-left-and-right
\let\originalright\right
\renewcommand{\left}{\mathopen{}\mathclose\bgroup\originalleft}
\renewcommand{\right}{\aftergroup\egroup\originalright}

\title{Partial differential equations to determine elasto-plastic stress-strain behavior from measured kinematic fields}

\author{
 Benjamin C. Cameron\\
  Department of Civil and Environmental Engineering\\
  Massachusetts Institute of Technology\\
  and Department of Mechanical Engineering\\
  University of Southampton\\
  email: \texttt{b.c.cameron@soton.ac.uk}
  \AND
   C. Cem. Tasan\\
   Department of Materials Science and Engineering\\
   Massachusetts Institute of Technology\\
   email: \texttt{tasan@mit.edu} \\
}

\begin{document}

\maketitle
\begin{abstract}
\begin{singlespace}
A system of partial differential equations (PDEs) is derived to compute the full-field stress from an observed kinematic field when the flow rule governing the plastic deformation is unknown. These equations generalize previously proposed equations that assume pure plastic behavior without elasticity. A method to numerically solve these equations is also presented. In addition to force balance, the equations are derived from the elastic-plastic decomposition of the deformation gradient, the assumption of isotropy, and the assumption that the function mapping the elastic strain to stress is known. The system of equations can be directly applied to complex geometries, finite deformation, non-linear elasticity and plasticity, compressible materials, rate dependent materials, and a variety of hardening laws. This system of PDEs is non-linear and time dependent. Furthermore, it overcomes an important prior limitation: it can be directly applied to cases where some regions of a body are elastically deforming while others are elasto-plastically deforming. A two-dimensional case study of necking in a uniaxial tensile specimen is investigated to illustrate and validate the method. The governing equations are numerically solved using strain fields output from a finite element simulation and validated against this same simulation showing accurate results.
\end{singlespace}
\end{abstract}

\section{Introduction}
\label{sec:introduction}
The rapid increase in availability and quality of full-field kinematic data from experiment has created a major imbalance between the available strain and stress information. Full two- or three-dimensional strain fields can be obtained by digital image correlation (DIC) or digital volume correlation \cite{Pan2009,Kang2005,Yan2015}, benefiting from tomographic imaging techniques where needed \cite{Lenoir2007,Roux2008,Tudisco2015}. However, force or stress measurements are largely restricted to load cells, \textit{in-situ} stress transducers \cite{Harris1994}, photoelasticity approaches \cite{Ramesh2000}, or destructive measurements \cite{Vermeij2018}. This poses major limitations in the analysis of heterogeneous deformations, as in the case of necking, shear band formation, or high strain-rate deformation. The kinematics of the deformation provides only limited insights regarding material behavior or properties, whereas knowing both the strain and the corresponding stress fields would determine the constitutive relations that fully define the mechanical response. 

This motivates the study of the inverse problem: computing the stress from the strain when the constitutive equations remain unknown. There has been some success in solving this problem in cases with homogeneous constitutive equations and a functional form known in advance. However, the requirements of force balance, an assumed constitutive equation form, homogeneity, and matching the observed strain field generally overdetermines the problem such that no solution can be found. To address this, optimization approaches have been applied to minimize error in one or more of these constraints. There have been a wide range of methods that take this approach, including the virtual field method \cite{Grediac2002,Florentin2011,Pierron2012,Marek2017}, the equilibrium gap method \cite{Crouzeix2009,Florentin2010}, the constitutive equation gap method \cite{Florentin2011}, finite element model updating \cite{Rethore2010,Siddiqui2017,Viala2018}, and others (e.g. \cite{Gelin1994, Leygue2018,Paranjape2021,Goenezen2021,Flaschel2022,Joshi2022a,Langlois2022}). These approaches appear to be well suited to noisy or incomplete data sets, however, depending on the application, the errors can be significant (typically ranging between 1 and 50 \%) and the computational cost can be high. These challenges are exacerbated when the material properties are heterogeneous, as is often the case when considering microstructures, graded materials, composite materials, or other heterogeneities arising from manufacturing processes. 

An alternate approach has been proposed which allows the constitutive equation to be heterogeneous and imposes weaker requirements on its functional form \cite{Cameron2021,Liu2021}. This leads to a system of partial differential equations (PDEs) for the stress field with a deterministic solution that is consistent with specific constitutive assumptions, the observed strain field, and force balance. This avoids the computationally expensive optimization procedures used in alternative methods. The key assumption made in these prior works is that the Cauchy stress is coaxial with either the observed strain (in the case of elastic deformation) or strain rate (in the case of pure plastic or viscous deformation). Coaxiality means that the two tensors will share the same eigenvectors or principal directions. These assumptions and the derived system of PDEs are discussed more fully in \cref{sec:background}. The main idea is that one can (i) use an approach such as DIC to measure the full-field strain, (ii) compute the eigenvectors at each point, and (iii) compute the stress by solving the system of equations with specified boundary conditions. In post processing, one can then compute stress-strain paths for each material point, determine hardening-laws, dependencies on stress invariants, dependencies on strain rate, or dependencies other variables that change as a function of position (material composition, temperature, etc.).

While this approach appears promising, some practical challenges require further consideration, for example, the experimental strain field uncertainty may exceed an acceptable value, the assumed boundary conditions may deviate from the true ones, non-unique solutions may arise in some cases, and the real material behavior may deviate from simplified models. Hence, in order to develop this approach into a widely used engineering and scientific tool, further research is necessary. This would likely include noise mitigation strategies, data pre- and post-processing methodologies, improved numerical methods, and equation modifications to incorporate more complex material models.

This article is focused on extending the approach to more complex material models because the prior assumptions made prevent one from determining the stress in many common cases. Frequently, regions of a deforming body will be undergoing significant plastic deformation while other regions will remain in the elastic deformation regime. One can either make the assumption that the deformation is elastic, or that the material is fully plastic. However, even in the case where these assumptions are reasonable for the different regions respectively, is not possible to know what the regions are. Hence, one cannot apply either of the assumptions above to determine the eigenvectors required for the system of partial differential equations. One example of this is a necking tensile specimen subject to displacement controlled boundary conditions. Here, the regions outside the neck will elastically unload while the region inside the neck will undergo substantial plastic deformation (see \cref{sec:proofofprinciple}). The boundary between these regions will be unknown and dynamically changing, hence, one cannot utilize either simplification to obtain the stress eigenvectors.

%why is elasto-plasticity important, or the lack of material models a challenge? "to further motivate..." does not quite make sense.

%Despite the fact that the equations are theoretically valid in the case of isotropic plastic deformation when the elastic strain rate is negligible, it is still not possible to apply this approach in its current form to the majority of these cases. For many cases of plastic deformation other regions of the body will remain in the elastic deformation regime. It is not possible to know the region boundaries without additional assumptions or knowledge of the material.

In this article we present an extension of this deterministic approach that applies to fully elasto-plastic deformation of isotropic materials. This will address the practical challenge discussed above, but will also address cases where the elastic and plastic strains are comparable and neither can be neglected. While the derived equations do not incorporate anisotropic models such as kinematic hardening or crystal plasticity, the model significantly increases the capability of the approach, and the mathematical framework presented may provide a foundation upon which some of these more complex models can be built.

%To further motivate a more general elasto-plastic approach, we note that this approach may be 

%emphasize that obtaining $\bQ$ directly from the rigid plastic approximation is not practical even if the plastic strain is much larger than the elastic strain in all regions of plastic deformation. 

In order to arrive at a deterministic system of partial differential equations, we make the assumption that the elastic strain is a known function of stress (we do not assume the elastic strain is known, only that the function itself is known). This assumption appears to be practical as the elastic moduli are often easily obtained (see discussion in \cref{sec:rangeofapplicability}). Even with the assumption that the elastic strain is a known function of the stress, the problem of determining the eigenvectors from the observed kinematics is non-trivial. Both the elastic strain and stress remain unknown and must be determined as part of the solution. In \cref{sec:background,sec:theory} we give the theoretical background and derive the non-linear time dependent system of PDEs. After linearizing the equations and providing some discussion (\cref{sec:linearization} and \cref{sec:rangeofapplicability} respectively), we consider a case study of practical significance: necking in a two-dimensional uniaxial tensile specimen (\cref{sec:proofofprinciple}). In order to validate the system of equations and numerical method, we use strain fields obtained from a finite element simulation.

\section{Background}
\label{sec:background}
\subsection{Kinematic framework}
\label{ssec:preliminaries}
We will predominantly use a finite strain elasto-plasticity framework, however, we present some background in the context of infintesimal strain, as the ideas are clearer
\footnote{Notation: Bold upper-case letters to denote second order tensors ($\mathbf{A},\mathbf{B}$,...), bold lowercase letters to denote vectors ($\mathbf{a},\mathbf{b}$,...), unbold letters correspond to scalars ($a$, $b$, $...$),  and the superscript $\top$ denote the transpose ($\mathbf{A}^\top$, $\mathbf{a}^\top$, $...$). We use standard matrix notation, for example, we have $\mathbf{a}\cdot\mathbf{b} = \mathbf{a}^\top\mathbf{b}$ and $\mathbf{a}\otimes\mathbf{b} = \mathbf{a}\mathbf{b}^\top$. An over bar denotes functional dependence, e.g. $\mathbf{A} = \bar{\mathbf{A}}(\mathbf{b})$. We denote the first and second material time derivative of a second order tensor $\mathbf{\dot{A}}$ and $\mathbf{\ddot{A}}$ respectively. $\mathrm{Div}\mathbf{A}$ and $\mathrm{div}\mathbf{A}$ correspond to the divergence $\mathbf{A}$ with respect to the initial and deformed reference frame respectively. $|\mathbf{A}|=\sqrt{A_{ij}A_{ij}}$ gives the magnitude of $\mathbf{A}$ (Einstein summation notation is used). We denote the deviatoric part of a second order tensor $\mathbf{A}$ as $\mathbf{A_0}$, and the trace as $\mathrm{tr}(\mathbf{A})$. The spectral decomposition of a symmetric tensor $\mathbf{A}$ is written as $\mathbf{A} = \mathbf{Q_\bA\Lambda_\bA Q_\bA^\top}$ where $\mathbf{Q_\bA}$ has columns with the eigenvectors of $\mathbf{A}$ denoted $\{\bq_\bA^{(i)}|i=1,2,3\}$ (principal directions) and $\mathbf{\Lambda_A}$ is a diagonal matrix containing the eigenvalues $\{\lambda_\bA^i|i=1,2,3\}$ (principal values). We use $o(f(x))$ to denote the set of functions such that $o(g(x))/f(x)\rightarrow 0$ as $x\rightarrow 0$.}.
In the case of infintesimal deformation, the displacement of a point $\bx$ at time $t$ is denoted $\bu = \bar{\bu}(\bx,t)$. Note that, for simplicity, we often suppress the arguments $\bx$ and $t$. The strain is defined as:
\begin{equation}
\bE = (\mathrm{grad}(\bu)+\mathrm{grad}(\bu)^\top)/2.
\end{equation}
In the case of elasto-plasticity, we assume an additive strain decomposition:
\begin{equation}
\bE = \bE^e+\bE^p,
\label{eq:temp18}
\end{equation}
where $\bE^e$ is the elastic strain and $\bE^p$ is the plastic strain. 

In the case of finite strain, we utilize a framework introduced in \cite{Gurtin2005a}. The reader can refer to \cite[Chapters~91-101]{Gurtin2010} for more details. We have a smooth one to one mapping from the reference body $\mathbf{X}$ to the deformed body given by $\mathbf{x}=\bar{\bx}(\mathbf{X},t)$. This can be used to specify the deformation gradient:
\begin{equation}
\bF = \mathrm{Grad}(\bx),
\end{equation}
and the velocity gradient:
\begin{equation}
\mathbf{L} = \dot{\mathbf{F}}\mathbf{F^{-1}}.
\end{equation} 
This can be decomposed into symmetric and skew parts: 
\begin{equation}
    \bD = (\bL+\bL^\top)/2, \qquad \bW=(\bL-\bL^\top)/2.
\end{equation}
The left and right Cauchy-Green tensors are given by:
\begin{equation}
    \qquad \bB = \bF\bF^\top, \qquad \bC = \bF^\top\bF.
\end{equation}
For elasto-plastic deformation we take the standard Kröner--Lee multiplicative decomposition:
\begin{equation}
\bF = \bF^e\bF^p.
\end{equation}
We define the left and right elastic Cauchy-Green tensors:
\begin{equation}
    \bC^e = \bF^{e\top}\bF^{e}, \qquad \bB^e = \bF^e\bF^{e\top}. 
\end{equation}
Furthermore, we define the variables:
\begin{equation}
    \bL^e = \dot{\bF}^e\bF^{e-1},\qquad \bL^p = \dot{\bF}^p\bF^{p-1}.
\end{equation}
Here, the reader should note that, in general, under this framework, $\bL \neq \bL^e+\bL^p$. This is similar to \cite{Lee1969}. We also define the variables:
\begin{equation}
    \bD^e=(\bL^e+\bL^{e\top})/2, \qquad \bW^e=(\bL^e-\bL^{e\top})/2,
\end{equation}
\begin{equation}
    \bD^p=(\bL^p+\bL^{p\top})/2, \qquad \bW^p=(\bL^p-\bL^{p\top})/2.
\end{equation}

\subsection{Coaxiality}
\label{ssec:coaxiality}
The system of equations derived in this article are based on assumptions relating the eigenvectors, or principal directions, of the Cauchy stress \bT with the eigenvectors, or principal directions, of other kinematic variables. We have:
\begin{equation}
    \bT\bq_\bT^{(i)} = \sigma^{(i)}\bq_\bT^{(i)},
\end{equation}
where $\{\bq_\bT^{(i)}\}$ are the orthonormal eigenvectors of $\bT$, $\{\sigma^{(i)}\}$ are the eigenvalues (or principal stresses), and $i \in \{ 1,2,3 \}$. The spectral decomposition of \bT is given by:
\begin{equation}
    \bT = \sum_i\sigma^{(i)}\bq_\bT^{(i)}\bq_\bT^{(i)\top}.
\end{equation}
In some cases, the eigenvalues may repeat. Let $k$, such that $k\subset\{1,2,3\}$, index the eigenvalues that repeat, so we have $\{\sigma^{(k)}\}$. The corresponding eigenvectors, $\{\bq_\bT^{(k)}\}$, are not unique and any linear combination of $\{\bq_\bT^{(k)}\}$ is also an eigenvector.

When the eigenvectors are unique, coaxiality is defined as sharing the same set eigenvectors. Let $\bQ_\bT$ be a matrix of eigenvectors that diagonalize the stress \bT, and $\bLambda_\bT$ be the matrix of eigenvalues, such that $\bT = \bQ_\bT\bLambda_\bT\bQ^\top_\bT$. Similarly, $\bE^e$ can be expressed as $\bQ_\bE^e\bLambda_\bE^e\bQ_\bE^{e\top}$. If $\bQ_\bT=\bQ_\bE^e$, then $\bT$ and $\bE^e$ are coaxial.

We show that coaxiality arises between the stress and various kinematic variables in a number of isotropic material models \cite{Dunne1984}. However, it does not arise in cases of anisotropic material models such as crystal plasticity, or anisotropic hardening models such as kinematic hardening.

First consider the infinitesimal strain case where $\bE^e$ is a function of \bT and other scalar variables $\{\alpha_i\}$:
\begin{equation}
\bE^e = f(\bT,\{\alpha_i\}).
\label{eq:temp31}
\end{equation}
If the function is isotropic, this requires that:
\begin{equation}
\mathcal{Q}^\top\bE^e\mathcal{Q} = f(\mathcal{Q}^\top\bT\mathcal{Q},\{\alpha_i\}),
\label{eq:temp32}
\end{equation}
for any rotation $\mathcal{Q}$. Physically, this corresponds to rotating the material prior to deformation and observing that the response is the same. In addition to invariance under rotation, we require invariance under reflection. Mathematically, $\mathcal{Q}\in \mathrm{Orth}$, where $\mathrm{Orth}$ is the group of orthonormal matrices with positive or negative determinant. For simplicity, we assume both $\bE^e$ and $\bT$ both have unique eigenvectors. If these conditions are met, we can prove that $\bE^e$ will be coaxial with \bT. 

Inserting the spectral decompositions for 
\bT and $\bE^e$ into \cref{eq:temp31} gives:
\begin{equation}
\bQ^e_\bE\bLambda^e_\bE\bQ_\bE^{e\top} = f(\bQ_\bT\bLambda_\bT\bQ_\bT^\top,\{\alpha_i\}).  
\end{equation}
We now apply \cref{eq:temp32} using $\mathcal{Q}=\bQ_\bT^\top$, and noting that $\mathcal{Q}\mathcal{Q}^\top=\bI$ for any orthonormal matrix:
\begin{equation}
\begin{aligned}
    \bQ^\top_\bT\bQ^e_\bE\bLambda^e_\bE\bQ_\bE^{e\top}\bQ_\bT &= f(\bQ^\top_\bT\bQ_\bT\bLambda_\bT\bQ_\bT^\top\bQ_\bT,\{\alpha_i\}),\\
    & = f(\bLambda_\bT, \{\alpha_i\}).
\end{aligned}
\label{eq:temp33}
\end{equation}
We now consider a reflection through the ($x_2$,$x_3$) plane characterized by $\mathcal{Q} = \mathcal{Q}_{R23}$ which is the diagonal matrix $\mathrm{diag}(-1,1,1)$. If applied to a symmetric matrix $\bA$, such that $\bA\rightarrow\mathcal{Q}_{R23}^\top\bA\mathcal{Q}_{R23}$, then the signs of $A_{12}$ and $A_{13}$ are reversed while the diagonal components remain unchanged. Applying this transformation to \cref{eq:temp33} we have:
\begin{equation}
\begin{aligned}
    \mathcal{Q}_{R23}\bQ^\top_\bT\bQ^e_\bE\bLambda^e_\bE\bQ_\bE^{e\top}\bQ_\bT\mathcal{Q}_{R23}&=
    f(\mathcal{Q}_{R23}\bLambda_\bT\mathcal{Q}_{R23}, \{\alpha_i\}),\\
    & =f(\bLambda_\bT, \{\alpha_i\}).
\end{aligned}
\label{eq:temp34}
\end{equation}
This last equation follows from the fact that $\bLambda_\bT$ is diagonal. Comparing \cref{eq:temp33} and \cref{eq:temp34} gives:
\begin{equation}
    \mathcal{Q}_{R23}\bQ^\top_\bT\bQ^e_\bE\bLambda^e_\bE\bQ_\bE^{e\top}\bQ_\bT\mathcal{Q}_{R23} = \bQ^\top_\bT\bQ^e_\bE\bLambda^e_\bE\bQ_\bE^{e\top}\bQ_\bT.
\end{equation}
Hence, the matrix $\bQ^\top_\bT\bQ^e_\bE\bLambda^e_\bE \bQ_\bE^{e\top} \bQ_\bT$ must have components in the (1,2) and (1,3) positions that are zero. We can apply the same argument using a reflection in the ($x_1$,$x_2$) plane to show that the (2,3) component will also be zero. Hence, the matrix $\bQ^\top_\bT \bQ^e_\bE \bLambda^e_\bE \bQ_\bE^{e\top} \bQ_\bT$ is diagonal. 

Because $\bQ^\top_\bT \bQ^e_\bE \bLambda^e_\bE \bQ_\bE^{e\top}\bQ_\bT$ is diagonal and $\bLambda^e_\bE$ is also diagonal, it follows that $\bQ^\top_\bT\bQ^e_\bE$ is a permutation matrix, multiplied by one of the diagonal matricies $\mathrm{diag}(\pm 1, \pm 1, \pm 1)$. As the order of the eigenvectors and signs of the eigenvectors are arbitrary, we take both the permutation and diagonal matrices to be $\bI$ respectively. This gives $\bQ^\top_\bT\bQ^e_\bE = \bI$. From this we can conclude that $\bQ_\bT = \bQ^e_\bE$. Hence, $\bT$ is coaxial with $\bE^e$.

A simple example of where the above assumptions apply is linear isotropic elasticity. However, they also apply to some non-linear models, and models that depend on variables such as temperature or strain rate.

In the case that $\bT$ does not have unique eigenvectors, one can show that, given the above assumptions, $\bE^e$ shares the same subspace of non-unique eigenvectors. It is possible, given the above assumptions, for $\bE^e$ to have a subset of non-unique eigenvectors when \bT has unique ones. However, the authors are not aware of any material models for which this occurs, excepting the case where $\bE^e=0$ (e.g. gas flow or rigid plasticity).

The plastic strain rate $\dot{\bE}^p$ is commonly specified as a function of \bT, and other variables, using a flow rule:
\begin{equation}
    \dot{\bE}^p = f_2(\bT,\{\beta_i\}).
\end{equation}
We again assume that the function additionally depends only on scalar variables $\{\beta_i\}$. These scalar variables could include the accumulated plastic strain, the initial yield stress, and the hardening coefficient. We exclude tensor valued variables, such as the translation of the yield surface which arises in kinematic hardening, the orientations of slip systems that control crystal plasticity, or other tensor-valued variables relevant to anisotropic phenomena. The function can nominally depend on the elastic strain provided this is an isotropic function of stress. If this function is isotropic and invariant under reflections, we can apply the same argument as presented above to show that $\bT$ is coaxial with $\dot{\bE}^p$. 

A common example where these assumptions apply is von Mises plasticity theory with isotropic hardening. However, it also applies to other yield surface shapes (such as the Tresca yield surface), or other dependencies (such as rate dependence, pressure dependence or temperature dependence). Nevertheless, the isotropy restriction excludes numerous flow rules. Note that it is possible, provided the assumptions are valid, for $\dot{\bE}^p$ to have non-unique eigenvectors when $\bT$ has unique eigenvectors. For example, when the sample has not yielded and $\dot{\bE}^p=0$.

Now consider the corresponding arguments applied to the finite strain framework. Given the Kröner-Lee multiplicative decomposition $\bF = \bF^e\bF^p$, we have that, $\bF^p$ provides a mapping from the material configuration to the structural configuration, and $\bF^e$ provides a mapping structural configuration to the spatial configuration \cite{Gurtin2005a}. The elastic deformation can be characterized by $\bC^e$ in the structural reference configuration, or $\bB^e$ in the spatial configuration. The plastic flow is characterized by $\bD^p$, which is defined in the structural configuration. A common stress variable, defined in the structural configuration is the second elastic Piola-Kirchoff  stress:
\begin{equation}
    \bT^e = \mathrm{det}(\bF^e)\bF^{e-1}\bT\bF^{e-\top}.
    \label{eq:temp35}
\end{equation}
Under the framework considered, one commonly specifies the elasticity relationship, and the flow rule, in the structural reference frame:
\begin{subequations}
\begin{gather}
    \bC^e = f_3(\bT^e,\{\gamma_i\})\\
    \bD^p = f_4(\bT^e,\{\delta_i\}).\label{eq:temp39}
\end{gather}

\end{subequations}
If $\{\gamma_i\}$ and $\{\delta_i\}$ are scalar variables and these functions are isotropic and reflection invariant, $\bC^e$ and $\bD^p$ will be coaxial with $\bT^e$. Examples where these assumptions apply include isotropic hyper-elasticity, and large deformation von Mises plasticity (as presented in \cite{Gurtin2010}). Furthermore, the assumptions made regarding $f_4$ are also valid for many viscous deformation models, for example, a classical incompressible Newtonian fluid where $\bD^p =\bD$ and $\bT^e=\bT$. We note that if we assume zero plastic spin, and no external body force, $\bT^e$ is the power conjugate to $\bD^p$.

We show in \ref{sec:finitecoaxiality}, that these coaxiality relationships translate directly to the coaxiality of \bT with $\bB^e$ and $\bF^e\bD^p\bF^{e\top}$. This follows from the fact that the right principal directions of $\bF^e$ are the same as the eigenvectors of $\bD^p$, and the left principal directions are the same as the eigenvectors of $\bB^e$. The prior discussion regarding non-unique eigenvectors also applies to these cases.

We note that this isotropy and reflective invariance of the function is a sufficient requirement to ensure that \bT and $\bE^e$ share eigenvectors, but it is not a necessary one. There may be models where these tensors share eigenvectors but the function is not isotropic.

To summarize, given appropriate assumptions regarding the isotropy and reflective invariance of the material, we can conclude that $\bT$ will be coaxial with $\bE^e$ and $\dot{\bE}^p$ in the infintesimal deformation case, and $\bB^e$ and $\bF^e\bD^p\bF^{e\top}$ in the finite deformation case.

%The key challenge in determining the principal directions of $\bT$ in the case of elasto-plastic deformation is that one can only directly obtain $\bB$ and $\bD$ from the observed deformation. $\bB^e$, $\bF^e$ and $\bD^p$ are unknowns. This is not a problem in the simplified case of elastic deformation as one can simply take $\bB^e = \bB$. Similarly in the case of rigid-plastic deformation, one can take $\bF^e\bD^p\bF^{e\top} = \bD$.

\subsection{Governing equations for simplified cases}
\label{ssec:simplifiedequations}
Equations to determine the stress field for the simplified cases of isotropic elastic or pure plastic/viscous deformation are introduced, as presented in prior works  \cite{Cameron2021,Cameron2022a,Liu2021}, to give the reader improved intuition when considering the more complex elasto-plastic case. We use $\{\bq^{(i)}\}$, without the subscript, to denote the eigenvectors of $\bT$ and other coaxial variables.

There are two simplified cases where $\{\bq^{(i)}\}$ can be trivially determined as a function of position from experimentally measured kinematic fields \cite{Cameron2021,Liu2021}:
\begin{enumerate}
    \item Elastic case ($\bE=\bE^e$): where $\bT$ is coaxial with $\bE$. $\{\bq^{(i)}\}$ can be obtained as a function of position using the spectral decomposition $\bE = \sum_i\lambda_\bE^{(i)}\bq^{(i)}\bq^{(i)\top}$. Alternatively, the equivalent approximation can be made when $\bB=\bB^e$ in the finite strain case. 
    \item Pure plastic case ($\dot{\bE}^e<<\dot{\bE}^p$ and $\dot{\bE}\approx\dot{\bE}^p$): where $\bT$ is approximately coaxial with the measured strain rate $\dot{\bE}$ and $\{\bq^{(i)}\}$ can again be obtained using the spectral decomposition. Alternatively, the equivalent approximation can be made when $\bD\approx\bD^p$ in the finite strain case. Note that this approximation also applies to viscous or visco-plastic deformation when the elastic strain rate is negligible or zero.
\end{enumerate}
Either case leads to a system of variable coefficient PDEs for \bT with $\{\bq^{(i)}\}$ known \emph{a priori}:
\begin{subequations}
\begin{gather}
    \mathrm{div}{\mathbf{T}} + \mathbf{b} = \rho\ddot{ \mathbf{x}}
     ,\label{eq:systemQa}\\
    \bT = \sum_i\sigma^{(i)}\bq^{(i)}\bq^{(i)\top}
    \qquad \mathrm{in} \qquad \Omega, \label{eq:systemQb} \\
    \bd^{(k)\top}\bT\bn = \bd^{(k)\top}\bt
    \qquad \mathrm{on} \qquad \partial\Omega \label{eq:systemQc},
\end{gather}
\label{eq:systemQ}
\end{subequations}
where $\mathbf{b}$ is the body force, $\ddot{\mathbf{x}}$ is the acceleration, $\rho$ is the density, $\bt$ is the traction, $\Omega$ is the domain of interest, and $\{\bd^{(k)}\}$ are unit vectors discussed later in this subsection\footnote{We refer to the system of algebraic equations and PDEs, simply as a system of PDEs because the equations can be combined into an equivalent system of only PDEs using substitution. However, it is not useful to make this substitution here.}.
\cref{eq:systemQa} is force balance and \cref{eq:systemQb} constrains $\bT$ to have the eigenvectors $\{\bq^{(i)}\}$. \cref{eq:systemQc} specifies the traction boundary conditions as discussed below. Assuming $\{\bq^{(i)}\}$, $\ddot{\mathbf{x}}$, $\rho$, and $\mathbf{b}$ and the boundary conditions are known, one can solve the linear, variable coefficient second-order, system of hyperbolic PDEs for the stress. In the finite strain case, these equations must be solved in the spatial configuration.

\begin{figure}[ht!]
\centering
\includegraphics[width=1\textwidth]{{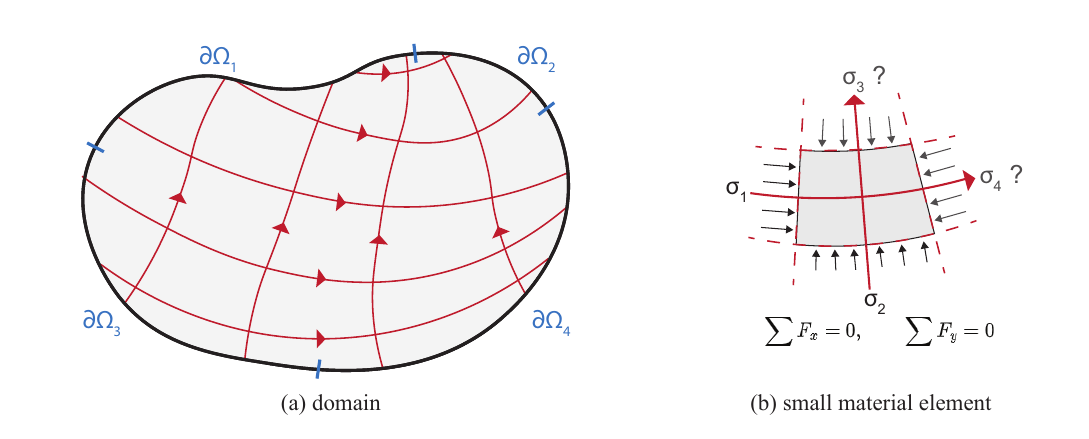}}
\caption{The structure of the hyperbolic system of PDEs when the eigenvectors $\{\bq^{(i)}\}$ are known. (a) A domain with characteristic curves that are aligned with the eigenvectors of the stress. These characteristic curves determine the direction of information propagation throughout the domain. Note that the characteristic curves are not necessarily parallel or perpendicular to the boundary. (b) A small material element with boundaries that are aligned with the characteristic curves. There can be no shear traction exerted across these boundaries. If $\sigma_1$ and $\sigma_2$ are known, there are only two unknowns: $\sigma_3$ and $\sigma_4$. These two unknowns can be determined using force balance. In this way, information propagates in the upward and rightward directions.}
\label{fig:overview}
\end{figure}

The structure of the system of PDEs is determined by the characteristic curves, which are aligned with the eigenvectors of stress at each point in the domain (\cref{fig:overview}). There are three sets of characteristic curves in three dimensions (two in two dimensions) and they will form a curvilinear system because the eigenvectors, and hence characteristic curves, are orthogonal (\cref{fig:overview}a). Consider a small material element with boundaries orthogonal to the characteristic curves (\cref{fig:overview}b). There can be no shear traction exerted on these boundaries, as the boundaries are oriented in the direction of the eigenvectors of stress. Hence, if we know the traction on half of the boundaries, we can directly calculate the traction on the remaining half using force balance. For the two dimensional case depicted in the figure, we can calculate the traction on the top and right hand side from the bottom and left side. Convergence of characteristic curves will amplify the stress since the cross-sectional area carrying the force is decreased, while curvature of the characteristics causes forces to be added/subtracted from perpendicular characteristics. If the traction is specified at the boundary where characteristics are entering the domain ($\partial \Omega_3$ in \cref{fig:overview}a), one can calculate the stress field in the upward and rightward directions by following the characteristic curves and using force balance as discussed above. This gives the system of equations its hyperbolic properties, as information starts on one boundary and is propagated in a specific direction throughout the domain, eventually determining the traction on other boundaries. Alternatively, one could specify different directions for information propagation and do the calculation in reverse. 

One cannot specify the traction boundary conditions for parts of the boundary where all characteristics are leaving the domain as they are calculated as part of the solution (e.g. $\partial \Omega_2$ in \cref{fig:overview}a). In cases where some of the characteristics are entering the domain and some are leaving ($\partial \Omega_1$ and $\partial \Omega_4$), one specifies the number of components of the traction equal to the number of characteristics entering the domain, with the condition that none of these components is directly aligned with characteristics leaving the domain. The directions in which the components of the traction are specified are $\{\bd^{(k)}\}$ in \cref{eq:systemQc} where $\{\bd^{(k)}\}$ will either span the whole space in the case where the traction is fully specified, span a subspace in cases were only a few components of traction can be specified, or may be the empty set in the case where the traction is determined as part of the solution.  For example, on $\partial \Omega_1$ and $\partial \Omega_4$, \textbf{$\{\bd^{(k)}\}$} will be a single vector chosen to be in any direction, except the direction aligned with the characteristic curve leaving the domain; on $\partial \Omega_3$, $\{\bd^{(k)}\}$ will be two vectors spanning the entire two dimensional space; and on $\partial \Omega_2$, $\{\bd^{(k)}\}$ will be the empty set.  This is derived and discussed in more detail in \cite{Cameron2022a}. Discontinuous stress fields can be addressed using the analysis presented in \cite{Cameron2022a}, which is important for composite materials or shock-wave phenomena.

In many cases the equations will give unique solutions, the primary exception arises in the case where the characteristic curves form closed loops. For example, consider a thick walled cylindrical pipe with a constant internal pressure causing expansion. As the problem is axi-symmetric, we would expect circular characteristic loops going around the circumference of the pipe. Other examples arise when a body with residual stresses has no traction on the boundaries. Of course, there are numerous scenarios where the equations give unique solutions: tensile experiments, bending, stress around crack tips, shear bands, shock waves, etc. Other issues solving the equations may arise when there are non-unique eigenvectors. This does not appear to be a major issue, because in the case of heterogeneous strain fields these repeated eigenvalues typically appear as singular points. One can exploit the global structure of the system of equations around these points to obtain a solution.

As mentioned in the introduction, this approach cannot be directly applied to elasto-plastic deformation because methods such as DIC will not give the elastic or plastic strain it will only give the total strain. The elastic-plastic strain decomposition will not be known and cannot be trivially determined, and hence the eigenvectors of the stress will not be known. Hence, a modified approach is required, which is the subject of this article.

\section{Governing equations for elasto-plasticity}
\label{sec:theory}

\subsection{Elasticity assumptions}
\label{ssec:assumption}
An assumption is made that the function $\bB^e = f(\bT)$ is known. To be clear, we do not assume $\bB^e$ is known, but only the functional form and associated parameters. The function may depend on other scalar variables, but these must be known. In the case of linear isotropic elasticity, this simply corresponds to knowing the elastic moduli. If the material is homogeneous, the elastic constants can be determined using standard techniques such as ultrasonic testing. Alternatively, if it is heterogeneous, the elastic constants can be computed from the initial elastic stage of deformation by solving \cref{eq:systemQ} while assuming isotropic elasticity (\cref{ssec:simplifiedequations}).

It is further assumed that $\bB^e$ has unique eigenvectors when \bT has unique eigenvectors. This is not required by the isotropy and reflective invariance of the function, discussed in \cref{ssec:coaxiality}. However, the assumption will almost always be true. Indeed, the authors are not aware of any isotropic elasticity models where this assumption can be violated, excepting models that assume no elastic strain (e.g. rigid plasticity or gas flow), and the assumption is not violated in the common cases of isotropic linear elasticity or hyper-elasticity. In cases where there is no elastic strain, one can solve the problem using the  pure plastic equations presented in \cref{ssec:simplifiedequations}.

\subsection{Derivation of governing equations}
\label{ssec:governingequations}
Here, we derive the elasto-plastic\footnote{We use the term "elasto-plastic" for simplicity, however, the equations will also be valid for elasto-visco-plastic deformation.} governing equations for the general case of finite deformation (\cref{eq:systemall}), however, we also present the simpler derivation for the infinitesimal case in \ref{sec:infintesimalderivation}. We encourage the reader to consult this first as the main lines of argument are clearer. We do not address the case of non-unique eigenvectors of \bT here but give further comments in \cref{ssec:repeatedeigen}.

Consider the deformation of a material element at times $t=\tau$ and $t=\tau+\Delta t$. We derive algebraic equations and then take the limit as $\Delta t \rightarrow 0$ to give the system of PDEs. We assume that the Cauchy stress $\bT$ is known at time $\tau$ and we proceed to determine $\bT$ at time $\tau+\Delta t$ based on the observed variable $\bF$ (\cref{fig:referenceframes}). We introduce subscripts to denote quantities evaluated at these times, e.g. $\bT_{\tau} = \bar{\bT}(\tau)$ and $\bT_{\tau+\Delta t} = \bar{\bT}(\tau+\Delta t)$. This increment of deformation between these two times is characterized by the deformation gradient $\mathbf{F}^*$:
\begin{figure}[ht!]
\centering
\includegraphics[width=0.75\textwidth]{{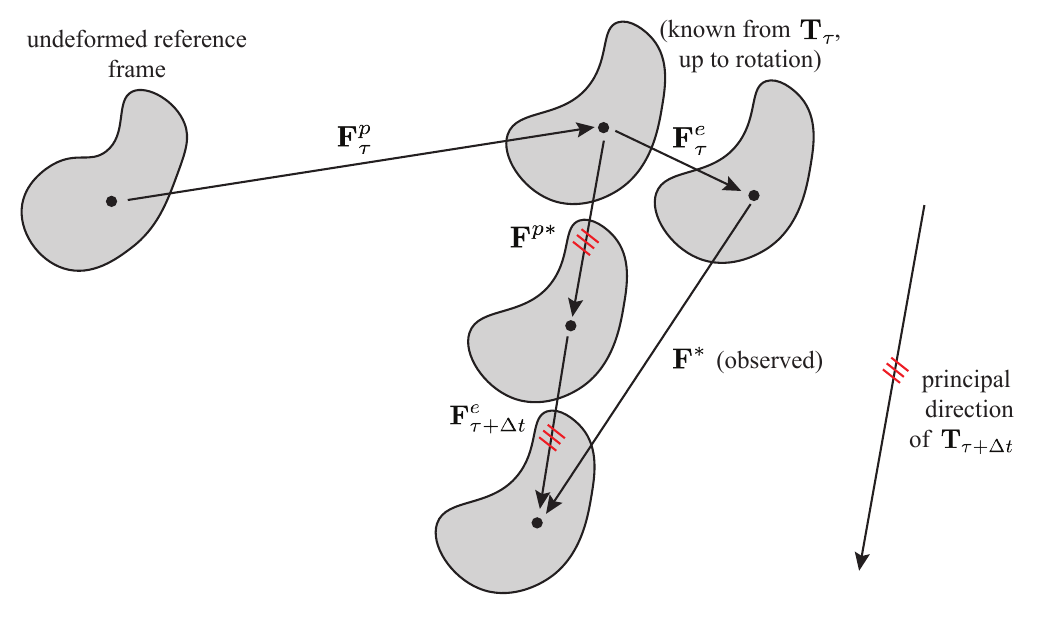}}
\caption{The references frames and deformation gradients used to map between them. The sample is initially deformed with $\bF_\tau = \bF_\tau^e\bF_\tau^p$, then an additional deformation $\bF^*$ is observed which is related to the unobserved $\bF^{p*}$ and $\bF^e_\tDt$. The triple red lines indicate that the eigenvectors can be directly related to one another. This is analogous to being parallel as the sum of two tensors with the same eigenvectors also have the same eigenvectors. }
\label{fig:referenceframes}
\end{figure}
\begin{equation}
    \bF_\tDt = \bF^*\bF_\tau. 
\end{equation}
We define $\bF^{p*}$ in the similar way:
\begin{equation}
    \bF^p_\tDt = \bF^{p*}\bF^p_\tau.
\end{equation}
$\bF^*$ and $\bF^{p*}$ are functions of both $\tau$ and $\tau+\Dt$. These variables tend to \bI as $\Dt\rightarrow 0$. We have:
\begin{equation}
    \bF^*\bF^{e}_{\tau}\bF^p_\tau = \bF^{e}_\tDt\bF^{p*}\bF^p_\tau,
    \label{eq:temp12}
\end{equation}
\begin{equation}
    \bF^*\bF^{e}_{\tau} = \bF^{e}_\tDt\bF^{p*}.
\end{equation}
This is the relationship depicted in the triangle in \cref{fig:referenceframes}. Note that the $\bF^e\bF^p$ decomposition is not unique due to an arbitrary rotation in the intermediate or structural reference frame \cite{Lee1969}. The following equations will be nevertheless be valid, and the final system of equations will be unique. Furthermore, variable combinations such as $\bF^e\bF^{e\top}$ and $\bF^e\bD^p\bF^{e\top}$ are unique and independent of the arbitrary rotation. 

Given the assumption that $\bB^e$ is coaxial with $\bT$ (\cref{ssec:coaxiality}), we have:
\begin{equation}
    \bB^e\bq^{(i)} = \bF^e\bF^{e\top}\bq^{(i)} = \lambda^{(i)} \bq^{(i)},
    \label{eq:temp8}
\end{equation}
where $\{\bq^{(i)}\}$ are the eigenvectors of both \bT and $\bB^e$, and $\{\lambda^{(i)}\}$ are the eigenvalues of $\bB^e$. Note that $\bB^e$ has unique eigenvectors, given the assumption discussed in \cref{ssec:assumption}, and that \bT is assumed to have unique eigenvectors.

\bT is coaxial with $\bF^e\bD^p\bF^{e\top}$ (\cref{ssec:coaxiality}). Hence, we have:
\begin{equation}
    \bF^e\bD^p\bF^{e\top}\bq^{(i)} = \gamma^{(i)} \bq^{(i)},
    \label{eq:temp9}
\end{equation}
where $\{\gamma^{(i)}\}$ are the eigenvalues and $\{\bq^{(i)}\}$ are the same eigenvectors of
\bT. This equation will be true in the case where $\bF^e\bD^p\bF^{e\top}$ does not have unique eigenvectors, provided the assumptions regarding isotropy and reflective invariance of the flow rule discussed in \cref{ssec:coaxiality} apply. This is important for cases when the material has not yielded and $\bD^p=0$, where $\{\gamma^{(i)}\}$ are simply all zero.

Consider the following expression:
\begin{equation}
    \bF^{e}_\tDt\bF^{e\top}_\tDt+
    2\Delta t\bF^{e}_\tDt\bD^p_\tDt\bF^{e\top}_\tDt.
\end{equation}
This will have the same eigenvectors as $\bT_\tDt$ because the sum of two tensors with the same eigenvectors also have the same eigenvectors and multiplication by a scalar preserves the eigenvectors. Furthermore, this quantity will have unique eigenvectors in the limit $\Dt\rightarrow 0$ provided $\bF^{e}\bF^{e\top}$ has unique eigenvectors as assumed above. 

Hence we have:
\begin{equation}
    \left(\bF^{e}_\tDt\bF^{e\top}_\tDt+
    2\Delta t\bF^{e}_\tDt\bD^p_\tDt\bF^{e\top}_\tDt\right)\bq^{(i)}_\tDt = ( \lambda_{\tDt}^{(i)}+2\Dt\gamma^{(i)}_\tDt)\bq^{(i)}_\tDt.
    \label{eq:temp38}
\end{equation}
We manipulate this in order to find an expression for the eigenvectors in terms of observed quantities. We have:
\begin{equation}
    \left(\bF^{e}_\tDt(\bI + 2\Delta t\bD^p_\tDt)\bF^{e\top}_\tDt\right)\bq^{(i)}_\tDt = ( \lambda_{\tDt}^{(i)}+2\Dt\gamma^{(i)}_\tDt)\bq^{(i)}_\tDt,
\end{equation}
\begin{equation}
    \left(\bF^{e}_\tDt(\bI + \Delta t\bL^p_\tDt+\Delta t\bL^{p\top}_\tDt)\bF^{e\top}_\tDt\right)\bq^{(i)}_\tDt = ( \lambda_{\tDt}^{(i)}+2\Dt\gamma^{(i)}_\tDt)\bq^{(i)}_\tDt.
    \label{eq:temp10}
\end{equation}
Note the following relationships:
\begin{equation}
    \begin{aligned}
    \bF^* &= \bF_\tDt\bF_\tau^{-1},\\
    &= \left(\bF_\tau +\Dt\dot{\bF}_\tau+ o(\Dt)\right)\bF_\tau^{-1},\\
    &= \bI + \Dt\bL_\tau+o(\Dt),
    \end{aligned}
    \label{eq:temp36}
\end{equation}

\begin{equation}
\begin{aligned}
    \bF^{p*} &= \bF^p_{\tDt}\bF^{p-1}_\tau,\\
    &=\left( \bF^p_\tau +\dot{\bF}^p_\tau \Dt+o(\Dt) \right)\bF^{p-1}_\tau,\\
    &= \bF^p_\tau\bF^{p-1}_\tau +\dot{\bF}^p_\tau\bF^{p-1}_\tau \Dt+o(\Dt) ,\\
    &=\bI + \Dt \bL^p_\tau + o(\Dt),\\
    &=\bI + \Dt \left( \bL^p_\tDt -\Dt \dot{\bL}^p_\tau +o(\Dt)\right) + o(\Dt)\\
    &=\bI + \Dt \bL^p_\tDt + o(\Dt).
\end{aligned}
\end{equation}
From this last expression we obtain:
\begin{equation}
    \begin{aligned}
        \bF^{p*}\bF^{p*\top} &= \left(\bI +\Dt\bL^p_\tDt+o(\Dt)\right)\left(\bI +\Dt\bL^{p\top}_\tDt+o(\Dt)\right),\\
        &=\bI + \Dt\bL^p_\tDt+ \Dt\bL^{p\top}_\tDt+o(\Dt).
    \end{aligned}
    \label{eq:temp11}
\end{equation}

Substituting \cref{eq:temp11} into \cref{eq:temp10} gives:
\begin{equation}
    \left(\bF^{e}_\tDt\bF^{p*}\bF^{p*\top}\bF^{e\top}_\tDt\right)\bq^{(i)}_\tDt +o(\Delta t) = ( \lambda_{\tDt}^{(i)}+2\Dt\gamma^{(i)}_\tDt)\bq^{(i)}_\tDt.
\end{equation}
From \cref{eq:temp12} we have $\bF^e_\tDt\bF^{p*} = \bF^{*}\bF^e_\tau$, hence:
\begin{equation}
    \left(\bF^{*}\bF^e_\tau\bF^{e\top}_\tau\bF^{*\top}\right)\bq^{(i)}_\tDt +o(\Delta t) = ( \lambda_{\tDt}^{(i)}+2\Dt\gamma^{(i)}_\tDt)\bq^{(i)}_\tDt.
    \label{eq:temp13}
\end{equation}

Provided $\Dt$ is small, this gives the eigenvectors of $\bT_\tDt$ in terms of the observed quantity $\bF^*$ and the known quantity $\bF^e_\tau\bF^{e\top}_\tau = \bB^e_\tau$. The latter of which can be computed from $\bT_\tau$ using the assumption that the elasticity equation is known (\cref{ssec:assumption}). This expression is used to derive the evolution equation for $\{\bq^{(i)}\}$. 

Substituting in \cref{eq:temp36} gives:
\begin{equation}
    (\bI + \Dt\bL_\tau+o(\Dt))\bF^e_\tau\bF^{e\top}_\tau(\bI + \Dt\bL^\top_\tau+o(\Dt))\bq^{(i)}_\tDt +o(\Delta t) = ( \lambda_{\tDt}^{(i)}+2\Dt\gamma^{(i)}_\tDt)\bq^{(i)}_\tDt,
\end{equation}
\begin{equation}
    \bF^e_\tau\bF^{e\top}_\tau\bq^{(i)}_\tDt+\Dt\left(\bL_\tau\bF^e_\tau\bF^{e\top}_\tau + \bF^e_\tau\bF^{e\top}_\tau\bL_\tau^\top\right)\bq^{(i)}_\tDt + o(\Dt)= ( \lambda_{\tDt}^{(i)}+2\Dt\gamma^{(i)}_\tDt)\bq^{(i)}_\tDt.
\end{equation}
From this, we subtract \cref{eq:temp8}, evaluated at $t=\tau$, giving:
\begin{equation}
    \begin{aligned}
    \bF^e_\tau\bF^{e\top}_\tau(\bq^{(i)}_\tDt-\bq^{(i)}_\tau)+\Dt\left(\bL_\tau\bF^e_\tau\bF^{e\top}_\tau + \bF^e_\tau\bF^{e\top}_\tau\bL_\tau^\top\right)\bq^{(i)}_\tDt + o(\Dt) \\ = (\lambda^{(i)}_\tDt\bq^{(i)}_\tDt - \lambda^{(i)}_\tau\bq^{(i)}_\tau)+2\Dt\gamma^{(i)}_\tDt\bq^{(i)}_\tDt.
    \end{aligned}
\end{equation}
Dividing by $\Dt$ and taking the limit as $\Dt
\rightarrow 0$ gives:
\begin{equation}
    \bF^e\bF^{e\top}\dot{\bq}^{(i)}+\left(\bL \bF^e\bF^{e\top} + \bF^e\bF^{e\top}\bL^\top\right)\bq^{(i)} = \frac{d}{dt}(\lambda^{(i)}\bq^{(i)}) +2\gamma^{(i)}\bq^{(i)},
\end{equation}
\begin{equation}
    \bB^e\dot{\bq}^{(i)}+\left(\bL \bB^e + \bB^e\bL^\top\right)\bq^{(i)} =
    (\dot{\lambda}^{(i)}+2\gamma^{(i)})\bq^{(i)}+
    \lambda^{(i)}\dot{\bq}^{(i)},
\end{equation}
where we have dropped the subscripts as all variables will simply depend on time $t=\tau$. We now multiply on the left side by $\bq^{(j)\top}$ for $j\neq i$, noting that $\bq^{(j)\top}\bq^{(i)}=0$ as the eigenvectors are orthogonal:
\begin{equation}
    \bq^{(j)\top}\bB^e\dot{\bq}^{(i)}+\bq^{(j)\top}\left(\bL \bB^e + \bB^e\bL^\top\right)\bq^{(i)} =
   \lambda^{(i)}\bq^{(j)\top}\dot{\bq}^{(i)},
\end{equation}
\begin{equation}
     \bq^{(j)\top}\left(\bB^e-\lambda^{(i)}\bI\right)\dot{\bq}^{(i)}+\bq^{(j)\top}\left(\bL \bB^e + \bB^e\bL^\top\right)\bq^{(i)} =
   0.
\end{equation}
Substituting in $\bq^{(j)\top}\bB^e = \lambda^{(j)}\bq^{(j)\top}$ and rearranging gives:
\begin{equation}
    \bq^{(j)\top}\dot{\bq}^{(i)} = 
    \frac{\bq^{(j)\top}\left(\bL \bB^e + \bB^e\bL^\top\right)\bq^{(i)}}{\lambda^{(i)}-\lambda^{(j)}},
\end{equation}
for all $j$ such that $j\neq i$.
Recall that we have made the assumption that the eigenvectors of $\bB^e$ are unique, hence $\lambda^{(i)}\neq\lambda^{(j)}$ for $i \neq j$. Since, $\bq^{(j)\top}\bq^{(j)}=1$
 and $\bq^{(j)\top}\bq^{(i)}=0$ for all $j$ such that $j\neq i$, we have:
\begin{equation}
    \dot{\bq}^{(i)} = \sum_{j,j\neq i}
    \frac{\bq^{(j)}\bq^{(j)\top}\left(\bL \bB^e + \bB^e\bL^\top\right)\bq^{(i)}}{\lambda^{(i)}-\lambda^{(j)}}.
    \label{eq:qdotBeL}
\end{equation}

We manipulate this expression further to give an alternate equation in terms of the velocity gradient $\bD$ and spin $\bW$. Using \cref{eq:temp8} we have:
\begin{equation}
\begin{aligned}
    \mathbf{q}^{(j)\top}(\bL\bB^e+\bB^e\bL^\top)\mathbf{q^{(i)}} &= 
    \lambda^{(i)}\bq^{(j)\top}\bL\bq^{(i)} +
    \lambda^{(j)}\bq^{(j)\top}\bL^\top\bq^{(i)},\\
    & = \bq^{(j)\top}(\lambda^{(i)}\bL+\lambda^{e(j)}_{\bB }\bL^\top)\bq^{(i)},\\
    & = \frac{(\lambda^{(i)}+\lambda^{(j)})}{2}\bq^{(j)\top}(\bL+\bL^\top)\bq^{(i)} +
    \frac{(\lambda^{(i)}-\lambda^{(j)})}{2}\bq^{(j)\top}(\bL-\bL^\top)\bq^{(i)},\\
    & =
    (\lambda^{(i)}+\lambda^{(j)})\bq^{(j)\top}\bD\bq^{(i)} +
    (\lambda^{(j)}-\lambda^{(i)})\bq^{(j)\top}\bW\bq^{(i)}.
\end{aligned}
\end{equation}
Substituting back into \cref{eq:qdotBeL} gives:
\begin{equation}
    \dot{\bq}^{(i)} = \sum_{j,j\neq i} \frac{\bq^{(j)}\bq^{(j)\top}\bD\bq^{(i)}(\lambda^{(j)}+\lambda^{(i)})}{\lambda^{(i)}-\lambda^{(j)}}
    + \sum_{j,j\neq i} \bq^{(j)}\bq^{(j)\top}\bW\bq^{(i)}.
    \label{eq:temp16}
\end{equation}
Note that for $i=j$, $\bq^{(j)\top}\bW\bq^{(i)} = 0$ as $\bW$ is skew. Hence, we can replace the second sum in \cref{eq:temp16} with a sum over all $j$. Then the term $\sum_{j}\mathbf{q}^{(j)}\mathbf{q}^{(j)\top}$ becomes the identity $\bI$. This gives:
\begin{equation}
\label{eq:qdotDW}
    \dot{\bq}^{(i)} = \sum_{j,j\neq i} \frac{\bq^{(j)}\bq^{(j)\top}\bD\bq^{(i)}(\lambda^{(j)}+\lambda^{(i)})}{\lambda^{(i)}-\lambda^{(j)}}
    +\bW\bq^{(i)}.
\end{equation}

When combined with force equilibrium and other algebraic equations reflecting the stated assumptions, this gives a complete set of PDEs to determine the stress $\bT$:
\begin{subequations}
\begin{empheq}[box=\widefbox]{gather}
    \mathrm{div}{\mathbf{T}} + \mathbf{b} = \rho\ddot{ \mathbf{x}},
        \label{eq:system0force}\\
        \dot{\bq}^{(i)} = \sum_{j,j\neq i} \frac{\bq^{(j)}\bq^{(j)\top}\bD\bq^{(i)}(\lambda^{(j)}+\lambda^{(i)})}{\lambda^{(i)}-\lambda^{(j)}}
            +\bW\bq^{(i)},
        \label{eq:system1qdot}\\
        \bT = \sum_i\sigma^{(i)}\bq^{(i)}\bq^{(i)\top}, \qquad \bB^e = \sum_i\lambda^{(i)}\bq^{(i)}\bq^{(i)\top},
         \label{eq:system2spectral}\\
         \bB^e = f(\bT), \qquad \mathrm{in} \qquad \Omega\\
            \label{eq:system3Be}
        \bd^{(k)\top}\bT\bn = \bd^{(k)\top}\bt
    \qquad \mathrm{on} \qquad \partial\Omega\\
    \bq^{(i)} = \bq_{t0}^{(i)} \qquad \mathrm{at} \qquad t=t_0.
\end{empheq}
\label{eq:systemall}
\end{subequations}
where $\Omega$ is the spatial domain, $\partial \Omega$ is the spatial domain boundary, and $\{\bq^{(i)}_{t0}\}$ are the principal directions at the initial time $t=t_0$. This system of equations only requires the observed variables $\bD$, $\bW$ and $\ddot{\bx}$ from the deformation, and other known variables such as the body force $\mathbf{b}$ and density $\rho$ in the case of dynamic deformations, and a subset of the full traction information on the boundary. If $\rho$ changes significantly throughout the deformation, it can be computed using the mass conservation equation with the measured kinematic data, provided the initial density distribution is known. The boundary conditions come from the same line of reasoning as \cite{Cameron2022a} which is briefly discussed in \cref{ssec:simplifiedequations}. Initially, one is only required to know $\{\bq^{(i)}\}$ because \bT($t_0$) can be determined by solving \cref{eq:system0force}. There are two major differences between this system of equations and \cref{eq:systemQ}. First, this system of equations is non-linear (due to \cref{eq:system1qdot}). Second, this system of equations is time dependent.

Existence and uniqueness proofs are challenging for non-linear systems of PDEs, and the problem remains unresolved for many established systems. Hence, rigorous analysis of this issue is outside the scope of the paper. Nevertheless, similarly to the simplified cases of elastic or pure plastic deformation, the solution will be non-unique when the characteristic curves form closed loops. Nevertheless, we note that solutions appear to be unique for many practical cases and further comments are given in \cref{sec:rangeofapplicability}.

In the case of linear elasticity where the elastic strains are small and the elastic constants are known, \cref{eq:system3Be} can be replaced with:
\begin{equation}
\label{eq:linelasticBe}
    \bE^e \approx \frac{1}{2}(\bB^e-\bI) = \frac{1}{2\mu}\left( \bT - \frac{\lambda}{2\mu+3\lambda} \mathrm{tr}(\bT)\mathbf{I}\right).
\end{equation}

\subsection{Non-unique eigenvectors}
\label{ssec:repeatedeigen}
Consider the case where $\bT$ does not have unique eigenvectors. Here, two or three of the eigenvalues repeat for both $\bT$ and $\bB^e$ (\cref{ssec:coaxiality}), the assumptions made in \cref{ssec:governingequations} are not be valid, and \cref{eq:system1qdot} is singular.

The simplest case is when \bT is homogeneous throughout space. Here, we can make an arbitrary choice for $\{\bq^{(i)}\}$, and use this over all space, provided we make the same choice for all $\bx$. An example of this is the trivial case when $\bT=0$ everywhere and there is no traction on the boundary. 

In the case where \bT is continuously varying and sufficiently differentiable, we can replace \cref{eq:system1qdot} with an equation that enforces continuity of the eigenvectors, or the derivatives of the eigenvectors throughout space and time \cite{Ojalvo1986,Friswell1996}. Here, the eigenvectors will only repeat in lower-dimensional subsets of the spatial and temporal domain. One then replaces \cref{eq:system1qdot} with the requirement that $\{\bq^{(i)}\}$ must be continuous across these subsets (or the derivatives must be continuous). In the spatial case, this requires detailed consideration of the global structure of the characteristic curves around these lower-dimensional subsets. While this is a critical topic, it is relatively involved and is hence outside the scope of this article. However, we do present here an important example of spatially homogeneous non-unique eigenvectors and how one can enforce the continuity of the eigenvectors with respect to time. Specifically, we consider stress free initial conditions in the next subsection.

\subsection{Initial conditions}
At time $t=0$ with stress free initial conditions, the choice of $\{\bq^{(i)}\}$ is non-unique. To resolve this ambiguity we require that $\{\bq^{(i)}\}$ are continuously varying and differentiable with respect to time. This gives rise to unique $\{\bq_{t0}^{(i)}\}$ \cite{Ojalvo1986,Friswell1996}. The subscript $t0$ is used to denote variables at time $t=0$. In the case of stress free initial conditions, we specify that $\{\bq_{t0}^{(i)}\}$ are equal to the eigenvectors at infinitesimal later time $t=\Delta t$, $\{\bq_{\Delta t}^{(i)}\}$.

First, we have that $\{\bq^{(i)}_{t0}\}$ are the eigenvectors of $\bB^e_\Dt$. The Taylor expansion for $\bB^e_\Dt$ is given by:
\begin{equation}
    \bB^e_\Dt = \bB^e_{t0}+\dot{\bB}^e_{t0}\Dt +o(\Dt).
\end{equation}
However, $\bB_{t0}=\bI$ because the material is unstressed at time $t=0$. Therefore, $\{\bq^{(i)}_{t0}\}$ are the eigenvectors of $\dot{\bB}^e_{t0}$. We have:
\begin{equation}
\begin{aligned}
    \dot{\bB}_{t0}^e &= \bL_{t0}^e\bB_{t0}^e+\bB_{t0}^e\bL_{t0}^{e\top},\\
    &=\bL_{t0}^e+\bL_{t0}^{e\top}.
\end{aligned}
\end{equation}
So $\{\bq^{(i)}_{t0}\}$ are the eigenvectors of $\bL_{t0}^e+\bL_{t0}^{e\top}$.

Second, we have that $\{\bq^{(i)}_{t0}\}$ are the eigenvectors of $\bF_{t0}^e\bD_{t0}^p\bF_{t0}^{e\top}$. However, as $\bF_{t0}^e = \bI$, the stress will be coaxial with  $\bD^p$. Given the definition of $\bD^p$, $\{\bq^{(i)}_{t0}\}$ will be the eigenvectors of $\bL_{t0}^p+\bL_{t0}^{p\top}$. 

As $\{\bq^{(i)}_{t0}\}$ are eigenvectors of both $(\bL_{t0}^e+\bL_{t0}^{e\top})$ and $(\bL_{t0}^p+\bL_{t0}^{p\top})$, we have that $\{\bq^{(i)}_{t0}\}$ must be the eigenvectors of $\bD_{t0} = \bL_{t0}+\bL_{t0}^\top$ given:
\begin{equation}
\begin{aligned}
    \bL_{t0} &= \bL_{t0}^e+\bF_{t0}^e\bL_{t0}^p\bF_{t0}^{e-1},\\
    &= \bL_{t0}^e+\bL_{t0}^p.
\end{aligned}
\end{equation}
Hence we have:
\begin{equation}
    \bD_{t0}\bq_{t0}^{(i)} = \lambda_\bD^{(i)}\bq_{t0}^{(i)}.
\end{equation}
Therefore, the initial $\{\bq^{(i)}\}$ can be
obtained from the observed quantity $\bD_{t0}$.

To summarize, if the material is initially in a stress-free state, one can specify $\{\bq_{t0}^{(i)}\}$ to be the eigenvectors of $\bD$.  If the material is not initially stress free, $\{\bq_{t0}^{(i)}\}$ must be specified using alternative assumptions or measurements.

\subsection{Limiting cases}
\label{ssec:limitingcases}
We consider several special cases and show how we arrive at the equations for small deformation, elastic deformation, and pure plastic deformation. 

First, consider the case of infinitesimal deformation. Here, $\bD \approx \bEd$. We also have that $\lambda^{(i)} $ is close to 1, expressed as  $\lambda^{(i)}  = 1 + 2\lambda^{e(i)}_{\bE}+o\left(\lambda_\bE^{e(i)}\right) = 1 + o(1)$ where $\{\lambda^{e(i)}_{\bE}\}$ are the eigenvalues of $\bE^e$. \cref{eq:qdotDW} becomes:
\begin{equation}
    \dot{\bq}^{(i)} = \sum_{j,j\neq i} \frac{\bq^{(j)}\bq^{(j)\top}\bEd\bq^{(i)}(2+o(1))}{2\lambda^{e(i)}_{\bE} - 2\lambda^{e(j)}_{\bE}+o\left(\lambda_\bE^{e(i)}\right) } + \bW\bq^{(i)}.
\end{equation}
We can neglect the $o(1)$ term in the numerator and the $o\left(\lambda_\bE^{e(i)}\right)$ term in the denominator. Because $\lambda^{e(i)}_{\bE}<<1$, the denominator is $o(1)$ and the second $\bW \bq^{(i)}$ term is negligible in comparison. This gives the following infintesimal strain formulation:
\begin{equation}
    \dot{\bq}^{(i)} = \sum_{j,j\neq i} \frac{\bq^{(j)}\bq^{(j)\top}\bEd\bq^{(i)}}{\lambda^{e(i)}_{\bE} - \lambda^{e(j)}_{\bE}}.
\end{equation}
Note this is derived in \ref{sec:infintesimalderivation} using a similar process to \cref{ssec:governingequations} but with an additive strain decomposition.

Second, consider the case of elastic deformation with no plastic deformation. Here, the system could be trivially solved by inverting $
\bB^e = f(\bT)$, however, we consider how the other equations in the system become equivalent to \cref{eq:systemQ} in the elastic limit. Here, $\bB^e = \bB$ and \cref{eq:qdotBeL} becomes: 
\begin{equation}
    \dot{\bq}^{(i)} = \sum_{j,j\neq i} \frac{\mathbf{q}^{(j)}\mathbf{q}^{(j)\top}\left(\bL\bB+\bB\bL^\top\right)\mathbf{q}^{(i)}}{\lambda^{(i)} - \lambda^{(j)}}.
\end{equation}
Note the following relationship:
\begin{equation}
    \dot{\bB} = \bL\bB+\bB\bL^\top.
\end{equation}
We now have:
\begin{equation}
    \dot{\bq}^{(i)} = \sum_{j,j\neq i} \frac{\bq^{(j)}\bq^{(j)\top}\dot{\bB}\bq^{(i)}}{\lambda^{(i)}_{\bB} - \lambda^{(j)}_{\bB}},
\end{equation}
where $\{\lambda_\bB^{(i)}\}$ are the eigenvalues of $\bB$. This is simply the expression for the rate of change of an eigenvector of $\bB$. I.e. for an arbitrary symmetric matrix $\bA$ which is smoothly changing in time, we have the following general expression for the derivatives of its eigenvectors $\{\bq_\bA^{(i)}\}$:
\begin{equation}
    \dot{\bq}^{(i)}_\bA = \sum_{j,j\neq i} \frac{\mathbf{q}_\bA^{(j)}\mathbf{q}_\bA^{(j)\top}\dot{\bA}\mathbf{q}_\bA^{(i)}}{\lambda^{(i)}_\bA - \lambda^{(j)}_\bA}.
    \label{eq:temp14}
\end{equation}
Hence, choosing $\bq$ to be an eigenvector of $\bB$ will solve the equation and we can replace \cref{eq:qdotDW} with:
\begin{equation}
    \bB \bq^{(i)} = \lambda_\bB^{(i)}{\bB}\bq^{(i)}.
\end{equation}
This directly reproduces the more simplified elastic case discussed in \cref{ssec:simplifiedequations}.

Third, consider the case of pure plastic deformation. We take the limit of \cref{eq:qdotDW} as $\bF^e \rightarrow \bI$ indicating no elastic strain. Here all $\lambda^{(i)} \rightarrow 1$ and the denominator of the first term on the right hand side approaches zero. Hence, in order to satisfy the equation in the limiting case, the numerator must approach 0. Hence, we have:

\begin{equation}
    \bq^{(j)\top}\bD\bq^{(i)} = 0 \qquad \textrm{for all}\quad i\neq j.
\end{equation}

Since the $ \bq^{(i)}$ vectors are orthogonal, i.e. $\bq^{(i)\top}\bq^{(j)}=0$ for all $i \neq j$, one can see that $\bq^{(i)}$ are the eigenvectors of $\bD$. This directly reproduces the result for the pure plastic case.

\section{Linearization}
\label{sec:linearization}
As the governing system of equations is non-linear, it is useful to derive the linearized form about some equilibrium solution \cite{Bathe1976}. Consider the linearization of \cref{eq:systemall}, about an equilibrium solution at time $t\sz$ during an arbitrary deformation. For simplicity, we assume quasi-static deformation with no body force, i.e. in the force balance equation we take $\ddot{\bx}= 0$ and $\mathbf{b} = 0$. Incorporation of these terms into the linearization can be carried out without significant challenges where needed. 

We introduce a small quantity $\epsilon$, where $\epsilon<<1$, to capture small changes from this equilibrium solution. We also use subscripts $[0]$ to denote equilibrium quantities and subscripts $[1]$ to denote deviations away from equilibrium:
\begin{subequations}
\begin{align}
    t &= t\sz + \epsilon t\so,\\
     \bq^{(i)} &= \bq^{(i)}\sz +\epsilon\bq^{(i)}\so+o(\epsilon) = \bq^{(i)}\sz +\epsilon\dot{\bq}^{(i)}t\so+o(\epsilon),
     \label{eq:linq}\\
    \sigma^{(i)} &= \sigma^{(i)}\sz+\epsilon \sigma^{(i)}\so+o(\epsilon).
    \label{linsigma}
\end{align}
\end{subequations}
In general we treat $\{\sigma^{(i)}\}$ and $\{\bq^{(i)}\}$ as our independent variables, though we will often refer to $\bT$ and $\bT\sz$ where convenient with:
\begin{equation}
     \bT\sz = \sum_i\sigma\sz^{(i)}\bq\sz^{(i)}\bq\sz^{(i)\top}.
\end{equation}
Furthermore, we simply refer to $\dot{\bF}$, $\{\dot{\bq}^{(i)}\}$ and $\bL$ with no subscripts as the second derivatives are be small in comparison to other terms:
\begin{equation}
    \bF = \bF\sz + \epsilon \dot{\bF}t\so+o(\epsilon).
\end{equation}
We also use $\bx\sz$, $\Omega\sz$ and $\partial \Omega\sz$ to denote the positions of material elements, the domain, and domain boundary at time $t\sz$. 

First, consider the equilibrium solution at some arbitrary stage in a deformation, where, by definition, the quantities are not changing with time. We assume that $\{\bq^{(i)}\sz\}$ is known. The quantities at time $t\sz$ will satisfy:
\begin{equation}
    \mathrm{div}\sz(\bT\sz) = 0, \qquad \mathrm{in}\qquad \Omega\sz,
\label{eq:linTeq}
\end{equation}
\begin{equation}
    \bd^{(k)\top}\bT\bn = \bd^{(k)\top}\bt\sz, \qquad \mathrm{on}\qquad \partial \Omega\sz,
\end{equation}
where $\mathrm{div}\sz$ is the divergence with respect $\bx\sz$, i.e. $\mathrm{div}\sz(\bA) = \partial A_{ij}/\partial x_{[0]j}$.

Next consider the linearized form of \cref{eq:system1qdot}. Substituting \cref{eq:qdotBeL} into \cref{eq:linq} gives:
\begin{equation}
    \bq^{(i)} = \bq^{(i)}\sz + \epsilon \bt\so \sum_{j, j \neq i} \frac{\bq\sz^{(j)}\bq^{(j)\top}\sz
    \left( \bL\bB^{e}\sz+\bB^{e}\sz\bL^{\top}\right)
    \bq\sz^{(i)}}{\lambda^{e(i)}_{\bB [0]} - \lambda^{e(j)}_{\bB [0]}}
    +o(\epsilon),
\end{equation}
where $\bB^e\sz = \bF\sz^e\bF\sz^{e\top}$ and $\{\lambda^{e(i)}_{\bB [0]}\}$ are the corresponding eigenvalues. We also have:
\begin{equation}
    \dot{\bq}^{(i)} = \sum_{j, j \neq i} \frac{\bq\sz^{(j)}\bq^{(j)\top}\sz
    \left( \bL\bB^{e}\sz+\bB^{e}\sz\bL^{\top}\right)
    \bq\sz^{(i)}}{\lambda^{e(i)}_{\bB [0]} - \lambda^{e(j)}_{\bB [0]}}
    +o(\epsilon),
    \label{eq:qlin}
\end{equation}
which gives the linearized form of \cref{eq:system1qdot}. 

Prior to deriving the linearized form of the force balance equation it is necessary to consider several relationships regarding the deformation gradient. We have:
\begin{equation}
    \bF = \bF\sz+\epsilon t\so\bL\bF\sz+o(\epsilon).
\end{equation}
We also define $\bF^* = \partial \bx/\partial \bx \sz$. We have:
\begin{subequations}
\begin{gather}
    \bF^* = \bF\bF^{-1}_{[0]},\\
    \bF^* = \bI + \epsilon t\so\bL + o(\epsilon) \label{eq:temp1_2}\\
    \bF^{*-1} = \bI - \epsilon t\so\bL + o(\epsilon),\\
    \mathrm{det}(\bF^*) = 1 + \epsilon t\so \mathrm{tr}(\bL) + o(\epsilon).
\end{gather}
\label{eq:temp1}
\end{subequations}
The referential form of the force balance equation gives:
\begin{equation}
    0 = \mathrm{div}\sz\left(\mathrm{det}(\bF^*)\bT \bF^{*-\top} \right),
\end{equation}
\begin{equation}
    0 = \mathrm{div}\sz\left(\mathrm{det}(\bF^*)\left(\sum_i\sigma^{(i)}\bq^{(i)}\bq^{(i)\top}\right) \bF^{*-\top} \right),
\end{equation}
\begin{equation}
    0 =  \mathrm{div}\sz\left( 
    \left(1+\epsilon t\so\mathrm{tr}(\bL)\right)
    \left(\sum_i\left(\sigma\sz^{(i)}+\epsilon\sigma\so^{(i)}\right)
    \left(\bq^{(i)}\sz+\epsilon\bq^{(i)}\so\right)
    \left(\bq^{(i)\top}\sz+\epsilon\bq^{(i)\top}\so\right)\right)
    \left(1-\epsilon t\so\bL^{\top}\right)
    \right) + o(\epsilon).
    \label{eq:temp17}
\end{equation}
Consider the summation component of this equation:
\begin{multline}
     \sum_i\left(\sigma\sz^{(i)}+\epsilon\sigma\so^{(i)}\right)
    \left(\bq^{(i)}\sz+\epsilon\bq^{(i)}\so\right)
    \left(\bq^{(i)\top}\sz+\epsilon\bq^{(i)\top}\so\right) \\ =  \sum_i\sigma^{(i)}\sz\bq^{(i)}\sz\bq^{(i)\top}\sz 
    +\epsilon\sum_i\sigma^{(i)}\so\bq^{(i)}\sz\bq^{(i)\top}\sz +
    \epsilon\sum_i\sigma^{(i)}\sz\bq^{(i)}\so\bq^{(i)\top}\sz +
    \epsilon\sum_i\sigma^{(i)}\sz\bq^{(i)}\sz\bq^{(i)\top}\so +
    o(\epsilon).
\end{multline}
We note that $\bq^{(i)}\so\bq^{(i)\top}\sz$ and $\bq^{(i)}\sz\bq^{(i)\top}\so$ will be  $o(1)$ since \cref{eq:qlin} requires $\dot{\bq}^{(i)}$ to be perpendicular to $\bq^{(i)}$ (with $o(1)$ error). Hence, this simplifies to:
\begin{equation}
     \sum_i\left(\sigma\sz^{(i)}+\epsilon\sigma\so^{(i)}\right)
    \left(\bq^{(i)}\sz+\epsilon\bq^{(i)}\so\right)
    \left(\bq^{(i)\top}\sz+\epsilon\bq^{(i)\top}\so\right), \\ =  \sum_i\sigma^{(i)}\sz\bq^{(i)}\sz\bq^{(i)\top}\sz 
    +\epsilon\sum_i\sigma^{(i)}\so\bq^{(i)}\sz\bq^{(i)\top}\sz +
    o(\epsilon).
\end{equation}
\cref{eq:temp17} becomes:
\begin{equation}
    \mathrm{div}\sz\left(
            \sum_i\sigma^{(i)} \bq\sz^{(i)}\bq\sz^{(i)\top}
        +\epsilon t\so\mathrm{tr}(\bL)\bT\sz
        -\epsilon t\so\bT\sz\bL^{\top}
    \right)
    + o(\epsilon)=0.
\end{equation}
Then substituting for $\epsilon t\so\bL$ using \cref{eq:temp1_2} we have:
\begin{equation}
   \mathrm{div}\sz\left(
            \sum_i\sigma^{(i)} \bq\sz^{(i)}\bq\sz^{(i)\top}
        +\mathrm{tr}(\bF^*-\bI)\bT\sz
        -\bT\sz\bF^{*\top}+\bT\sz
    \right)
    + o(\epsilon) = 0.
\end{equation}
Using \cref{eq:linTeq} gives:
\begin{equation}
    \mathrm{div}\sz\left(
            \sum_i\sigma^{(i)} \bq\sz^{(i)}\bq\sz^{(i)\top}
        +\mathrm{tr}(\bF^*)\bT\sz
        -\bT\sz\bF^{*\top}
    \right)
    + o(\epsilon) = 0.
\end{equation}
This gives the linearized form of force balance.

We denote $\bt^*$ as the first Piola-Kirchhoff traction vector with respect to the $\bx\sz$ reference frame such that:
\begin{equation}
    \mathrm{det}(\bF^*)\bT\bF^{*-\top}\bn\sz = \bt^*.
\end{equation}

Now consider the limit as $\epsilon \rightarrow 0$ to obtain the linearized system:
\begin{subequations}
\begin{gather}
    \mathrm{div}\sz\left(
            \sum_i\sigma^{(i)} \bq\sz^{(i)}\bq\sz^{(i)\top}
        +\mathrm{tr}(\bF^*)\bT\sz
        -\bT\sz\bF^{*\top}
    \right) = 0,\\
    \dot{\bq}^{(i)} = \sum_{j, j \neq i} \frac{\bq\sz^{(j)}\bq^{(j)\top}\sz
    \left( \bL\bB^{e}\sz+\bB^{e}\sz\bL^{\top}\right)
    \bq\sz^{(i)}}{\lambda^{e(i)}_{\bB [0]} - \lambda^{e(j)}_{\bB [0]}},\\
    \bT = \sum_i\sigma^{(i)}\bq^{(i)}\bq^{(i)\top} \qquad \textrm{in}\qquad \Omega\sz,\\
    \bd^{(k)\top}\mathrm{det}(\bF^*)\bT\bF^{*-\top}\bn\sz = \bd^{(k)\top}\bt^*
    \qquad \mathrm{on} \qquad \partial\Omega\sz.
\end{gather}
\end{subequations}

The linearized form corresponding to infinitesimal deformation is also important because in general, deformations where linearization is valid will be small. They can be obtained via the same procedure as above or by simplifying the above equations. This gives the following linearized system for small changes in $\bT$ and requiring small changes is $\{\bq^{(i)}\}$:
\begin{subequations}
\begin{gather}
   \mathrm{div}\left(
    \sum_i\sigma^{(i)} \bq\sz^{(i)}\bq\sz^{(i)\top}\right) = 0,\\
    \dot{\bq}^{(i)} = \sum_{j, j \neq i} \frac{\bq\sz^{(j)}\bq^{(j)\top}\sz
    \dot{\bE}
    \bq\sz^{(i)}}{\lambda^{e(i)}_{\bE [0]} - \lambda^{e(j)}_{\bE [0]}},\\
    \bT = \sum_i\sigma^{(i)}\bq^{(i)}\bq^{(i)\top} \qquad \textrm{in}\qquad \Omega,\\
    \bd^{(k)\top}\bT\bn = \bd^{(k)\top}\bt
    \qquad \mathrm{on} \qquad \partial\Omega.
\end{gather}
\end{subequations}

\section{Discussion}
\label{sec:rangeofapplicability}
The derived system of equations (\cref{eq:systemall}) allows one to compute the full-field stress from measured full-field kinematics and traction boundary conditions for elasto-plastic materials provided the stated assumptions are satisfied, when the flow rule governing the plastic deformation is unknown. This was previously only possible for isotropic elastic or pure plastic deformation (\cref{ssec:simplifiedequations}) \cite{Cameron2021,Liu2021}. However, there are practical considerations and issues that may arise when applying this approach to determine material properties. The assumptions made impose constraints on the systems that can be investigated, and there may be challenges acquiring the full-field kinematic data and measuring/estimating the boundary and initial conditions.

The simplified system of equations (\cref{ssec:simplifiedequations}), developed in prior works, cannot be applied to most heterogeneous plastic deformations because typically some regions will be elastic (\cref{sec:introduction}). The equations derived here resolve this issue, and also allow computation of the stress when both the elastic and plastic strain are significant. The primary assumption in this model is material isotropy, specifically, the isotropy of the elasticity relationship and flow rule, with respect to the structural reference frame, throughout the deformation. This is a weaker assumption than assuming specific isotropic functional forms, for example, von Mises plasticity and linear hardening. The system of equations can be applied to cases with more complex dependencies, for example rate dependence, temperature dependence, or pressure dependence. The functional forms given in \cref{ssec:coaxiality} can be used to determine what material models may be admissible. The only modification to the system of equations required when considering different material models is the known elasticity relationship. The most important limitation of the approach appears to be that it is restricted to isotropic material models and excludes important anisotropic cases such as textured metals, composite materials, crystal plasticity, and situations where kinematic hardening is significant. These may remain unresolved challenges, however, it may also be possible to address some of these cases in future work using alternate assumptions. 

In the case of a linear elasticity relationship, the assumption that this function is known (\cref{ssec:assumption}) is not likely to impose major practical restrictions, except in cases where the elastic moduli change throughout the deformation. If this is not the case, the elastic moduli can be measured in advance using conventional methods. If the elastic properties are heterogeneous and unchanging, they can be determined using the elastic system of equations discussed in \cref{ssec:simplifiedequations} during an initial elastic portion of the deformation. However, a practical challenge arises if the elastic moduli change throughout the deformation due to microscopic damage \cite{Yeh2003}, or other mechanisms \cite{Cleveland2002}. If the elastic constants are functions of changing scalar variables, such as temperature or chemical composition, the functional dependence must be known/assumed, and these variables determined, so that the elastic strain can be determined as a function of stress (\cref{eq:system3Be}).  For example, in the case of variable temperature, it may be possible to estimate the temperature distribution by modeling the temperature evolution (potentially using the kinematic data and computed stress). Alternatively, one may be able to measure the temperature throughout deformation using infrared thermography \cite{Wei2019}. Either approach may impose challenges, and application of the proposed methodology could be prevented in some cases.  We note that the derived system of equations is valid in the case of non-linear isotropic elasticity. However, it is unclear if there are scenarios where one can determine a non-linear elasticity relationship at high strains, but where it would still be useful to determine the flow rule using the proposed approach. The assumption that $\bB^e$ will have unique eigenvectors if \bT has unique eigenvectors, is not likely to restrict applications, as discussed in \cref{ssec:assumption}.

The derived equations will be directly applicable to some elasto-visco-plastic and visco-elastic deformation models, given that the coaxiality assumptions are directly applicable to isotropic models of viscous deformation (\cref{ssec:coaxiality}). However, many elasto-visco-plastic models are not based on the elasto-plastic Kröner decomposition, and are instead formulated using an evolution equation. This prevents the developed approach from being directly applied. However, these alternate formulations are closely related and can often be used to describe the same phenomena (\cite{Rubin1993} and references therein), so the approach presented may nevertheless be useful.

The traction boundary conditions must be known. The ability to satisfy this requirement is problem dependent, for example, when deforming a tensile sample, one may accurately assume that the stress is constant along the gauge, far away from the edge or neck. Furthermore, it is possible to  experimentally determine the boundary conditions in some cases. If the experiment is designed such that the material at the boundary is elastic, the elastic constitutive equation is known, and the strain is known, one can directly compute the stress and traction. Note that the normalized error in the boundary conditions will correspond to the order of magnitude of the normalized error in the stress field \cite{Cameron2021}. The initial conditions must also be known. In the common case of stress free initial conditions, this is trivial, however in other cases, it may be challenging or infeasible.

The existence and uniqueness of solutions to non-linear systems of partial differential equations is an involved subject, and a rigorous analysis is outside the scope of this article. However, some intuition can be gained by considering the variations in space and time separately. When solving the system numerically, if one computes an updated $\{\bq^{(i)}\}$ for a new time step, one can use this to solve for the stress at all positions using the approach in \cref{ssec:simplifiedequations}. Here, uniqueness is more readily understood, as it  corresponds to the characteristic curves not forming closed loops. We know, from the analysis of stress fields in known cases, that the characteristic curves will not form closed loops and give unique solutions in a large number of problems, for example, bending, necking, and shear band formation. However, the characteristic curves give non unique solution in examples such as axisymmetric expansion, or a traction free body with a residual stress field. Of course, there may be nonlinear couplings between $\{\bq^{(i)}\}$ and $\{\sigma^{(i)}\}$ that give rise to other existence or non-uniqueness issues.

The necessary full-field deformation measurements required to apply the approach in two dimensions can be obtained using DIC \cite{Pan2009}, the grid method \cite{Chalal2006,Moulart2011}, or particle image velocimetry \cite{Adrian2005}. \cref{sec:proofofprinciple} shows that the three dimensional system of equations can be reduced to two dimensions, under the assumption that the out of plane shear components are negligible, the assumption that the stress is homogeneous in the third dimension, and the assumption that the material is incompressible. Stereo-DIC \cite{Schmidt2003a} may be useful in future work for determining thickness of the sample in cases where this last assumption is not valid. However, the two dimensional system of equations cannot be applied to out of plane bending using this data.  In three dimensions, the kinematic fields can be determined using tomographic imaging techniques coupled with techniques such as digital volume correlation \cite{Lenoir2007,Roux2008,Tudisco2015}.

When considering the effect of experimental error $\bF'$ in the observed deformation gradient $\bF$, one must consider (i) the error introduced in $\{\bq^{(i)}\}$ and (ii)  how the error in $\{\bq^{(i)}\}$ translates to error in the stress. The latter involves non-local effects throughout the domain, and while this is quantified for one example in \cite{Cameron2021}, rigorous theoretical analysis is outside the scope of the present work. The relationship between the experimental error $\bF'$ and the error in $\{\bq^{(i)}\}$ is also complex, but can be partially understood by considering the elastic and pure plastic limiting cases of the system of equations. First, consider the elastic limit where $\{\bq^{(i)}\}$ are computed directly from $\bB^e+\bB^{e\prime}$, where $\bB^{e\prime}$ is the error. Assuming $\bB^{e\prime}<<\bB^e$, and given the theory of matrix perturbations, we have that the error in $\{\bq^{(i)}\}$ will scale with $|\bB_0^{e\prime}|$. The error will be small if $\bB_0^{e\prime}<<\bB_0^e$. Second, consider the plastic limit, where the eigenvectors are computed directly from $\bD^p +\bD^{p\prime}$ where $\bD^{p\prime}$ is the error. Again, assuming $\bD^{p\prime}<<\bD^p$, we have that the error in $\{\bq^{(i)}\}$ will scale with $|\bD_0^{p\prime}|$. The error will be small if $\bD_0^{\prime}<<\bD_0$. In intermediate cases with both elastic and plastic deformation, there will be a non-linear interaction of these errors. For example, consider plastic deformation followed by elastic unloading, the error in the stress computed from the plastic deformation will translate to an error in the stress during elastic unloading because the stress at the beginning of the unloading stage will be erroneous. Nevertheless, the errors in the elastic and plastic limits may prove useful guidelines until this non-linearity is quantitatively understood. To summarize, we require that:
\begin{equation}
    \begin{gathered}
    \bB_0^{e\prime}<<\bB_0^e,\\
    \bD_0^{p\prime}<<\bD_0^{p}.
    \end{gathered}
\end{equation}
The accuracy of typical DIC strain data typically ranges between $1\times10^{-3}$ and $1\times10^{-5}$. Hence, the approach may be viable given current levels of accuracy --- although this clearly needs investigation using experimental data and/or noisy synthetic detests. In addition to the error in the stress depending on the error in the strain/strain rate, it will depend on the strain and strain rate gradients, as well as the magnitude of these quantities. Depending on the noise level, it may be necessary to implement more specialized numerical discretizations, data smoothing methods, or other error mitigation techniques. For discussion and results related to the noise sensitivity of \cref{eq:systemQ}, the reader should refer to \cite{Cameron2021}.

\cref{eq:systemall} does not ensure thermodynamic consistency, although if the assumptions and data are correct throughout the deformation, the result will be thermodynamically consistent.  Consider a trivial counter example. A homogeneous cube is in quasi-static equilibrium and subject to spatially-uniform tensile traction on the faces orientated in the $x$ and $-x$ directions. The cube is observed to homogeneously compress along the $x$-direction. The equations can be solved to obtain a homogeneous tensile stress field. If the elastic constitutive law specified is thermodynamically admissible, the plastic stress-strain relationship will violate the Clausius-Duhem non-negative dissipation condition. This demonstrates that the system of equations may result in a stress field that is not thermodynamically consistent. Clearly, this hypothetical example will not occur for a standard material, but there may be more subtle thermodynamic violations because of violated assumptions, erroneous kinematic fields, or erroneous boundary conditions. A thermodynamic consistency check at every point could be applied during post-processing, to assess the quality of the data and assumptions.

Numerical solutions can be obtained using the algorithm outlined in \cref{ssec:inverseproblem}. This is based on the finite volume method developed in \cite{Cameron2022a}, and can be directly applied to general geometries and general boundary conditions.

The approach is not restricted to mechanically induced deformation, but applies also to thermally, chemically, or electrically induced deformation provided that the traction at the boundary (and/or body forces) are known, the kinematic field can be measured, and the constitutive assumptions are valid. 
 
\section{Case study: necking of a uniaxial tensile specimen}
\label{sec:proofofprinciple}
The governing equations are applied to a strain field output by a finite element simulation. This allows us to validate the equations by comparing the computed stress field to that output by the finite element simulation. We refer to the finite element simulation as the forward problem, and solving the derived system of equations as the inverse problem. We consider a tensile test conducted on a two-dimensional sample, inspired by typical tensile tests of ductile sheet metal where DIC data is readily available. In a real experiment, the sample will typically deform with a relatively constant cross-section, a diffuse neck will then develop, a localized neck will form, and the sample will fracture \cite{Ghosh1974}. As we only conduct a two-dimensional simulation, we exclude simulation of localized necking which is inherently three dimensional. Applying the equations and the numerical algorithm discussed below will demonstrate the approach when (i) the effects of finite deformation are significant, and (ii) there are both elasto-plastically and elastically deforming regions. Furthermore, the problem is of practical importance for obtaining the constitutive equations of a sheet metal at high strains.

There is existing literature on estimating the stress in a necking tensile specimen. Two geometries have been of particular interest: cylindrical specimens (for the analyses of thick plates or other bulk material), or flat specimens (for the analyses of sheet materials). The most well known approximation used for cylindrical specimens is the Bridgman solution which is based on the assumption of constant equivalent strain throughout the cross section \cite{Bridgman1952}. A common assumption used in in the case of flat specimens, which we adopt, is that the material is in a state of plane stress. This assumption will be valid during the formation of a diffuse neck but will break down during localized necking \cite{Ghosh1974}. The full-field strain can be measured for flat rectangular specimens using DIC \cite{Cameron2019b} and various error-minimization solutions have been applied to estimate the stress. Iterative finite element methods minimize the difference in assumed strain and measured strain (for example, \cite{Kajberg2004}). Alternatively, the virtual field method minimizes the error in the force balance equation \cite{Kim2013}. Other approaches minimize error in work applied to the neck (e.g. \cite{Coppieters2016}), minimize the error in the force-displacement curve measured for the sample \cite{Zhano1994}, or estimate and apply correction factors \cite{Zhang1999}. However, all of these approaches involve assumptions about the hardening behavior and do not give exact solutions. Errors vary depending on the material under investigation, the data available, and the strain level, but are typically lower than more general applications of the inverse problem discussed in the introduction, as the material properties are homogeneous (typically errors range between 1-10\%). For a more comprehensive review the reader should refer to \cite{Tu2019}.

The following methodology, explained in \cite{Cameron2021}, is implemented to validate the equations and numerical method. First, we run a forward problem computation using a commercial finite element solver which outputs the stress and displacement fields (\cref{ssec:forwardproblem}). Second, we export the full field kinematics to use as input to the inverse problem. We use this, along with the boundary conditions, to calculate the stress-field using \cref{eq:systemall} (\cref{ssec:inverseproblem}). Third, we compare the stress field output from the forward problem, with that output from the inverse problem, to assess the accuracy (\cref{ssec:results}). 

% We have a two dimensional specimen that is rectangular in the undeformed frame with dimensions $l_X$ and $l_Y$ in the $X$ and $Y$ directions. We have $l_X = 6l_Y$. $(X,Y)$ corresponds the the location in undeformed frame, and $(x,y)$ corresponds to the position in the deformed frame.  We have boundaries at $X=0$, $X=l_X$, $Y=0$ and $Y=l_Y$.
\subsection{The forward problem}
\label{ssec:forwardproblem}
\begin{figure}[ht!]
\centering
\includegraphics[width=1\textwidth]{{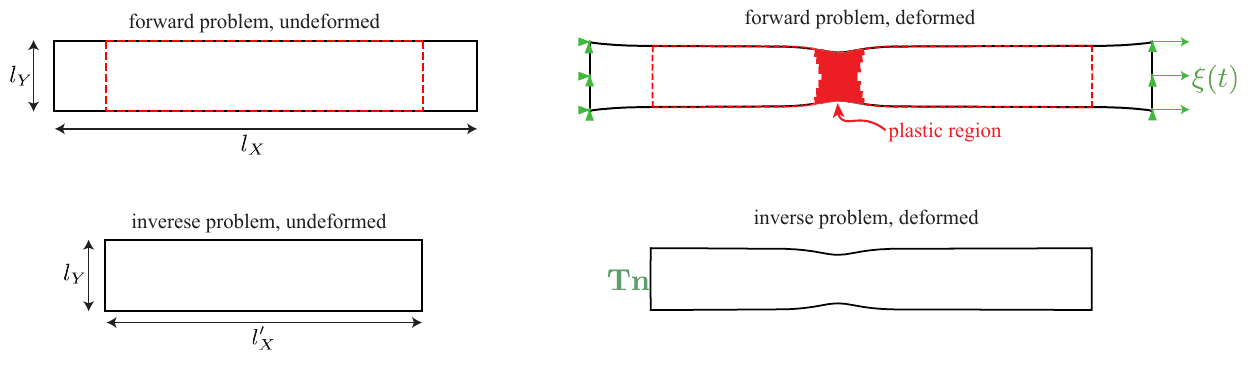}}
\caption{Domain and boundary conditions for forward and inverse problem. The inverse problem domain is chosen as a subset of the forward problem domain (highlighted in red). Green shows the boundary conditions: displacement for the forward problem and traction for the inverse problem. }
\label{fig:boundaryconditions}
\end{figure}

We have a two-dimensional specimen subject to plane stress that is rectangular in the undeformed frame with dimensions $l_X$ and $l_Y$ in the $X$ and $Y$ directions (\cref{fig:boundaryconditions}). We have $l_X = 6l_Y$. $(X,Y)$ corresponds the the position in undeformed frame, and $(x,y)$ corresponds to the position in the deformed frame.  We have boundaries at $X=0$, $X=l_X$, $Y=0$ and $Y=l_Y$. 

We specify the displacement on the $X=0$ and $X=l_x$ boundaries so that the distance between the boundaries increases as a function of time. To ensure that the sample necks somewhere near the center of the gauge, we specify that the $y$ displacement is zero on these boundaries. We specify zero traction on the $Y=0$ and $Y=l_Y$ boundaries.

We use the ABAQUS/explicit (2017) finite element solver for the forward problem and specify the following boundary conditions:

\begin{equation}
    \begin{gathered}
    x(X=0,Y) = 0, \qquad y(X=0,y) = 0,\\
    x(X=l_x,Y) = \xi(t), \qquad y(X=l_x,y) = 0,\\
    \bt(X,Y=0) = 0,\\
    \bt(X,Y=l_y) = 0.\\
    \end{gathered}
\end{equation}

\begin{figure}[!htb]
\centering
\includegraphics[width=1\textwidth]{{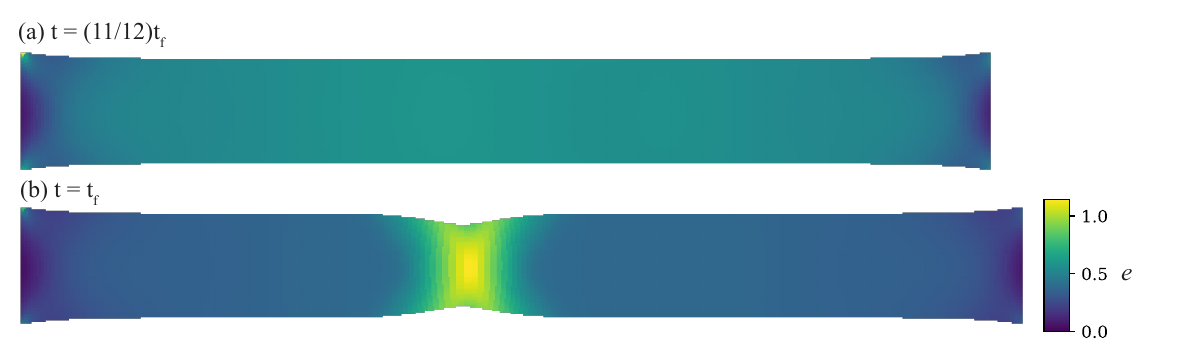}}
\caption{Strain output from forward problem at (a) $t=(11/12)t_f$ and (b) $t=t_f$. In this figure we use the scalar strain measure $e = \mathrm{ln}(1+\sqrt{2/3}|\bE|)$.}
\label{fig:strainoutput}
\end{figure}

Here, $\xi(t)$ is is a cubic function specified so that $d\xi(t=0)/dt = d^2\xi(t=0)/dt^2 = 0$ (this ensures zero acceleration at ${t=0}$). We specify a homogeneous elasto-plastic constitutive equation. Linear isotropic elasticity is used with a Young's Modulus ${E = 100}$~GPa and a Poisson's ratio of ${\nu = 0.3}$. For the plastic properties, we specify isotropic hardening with a von Mises yield surface and the yield stress as a piecewise linear function of the accumulated plastic strain, ${\varepsilon = \sqrt{2/3}\int_0^t|\bD^p|dt}$. This piecewise linear function is shown later in the article in \cref{fig:stressstrain}. More complex effects such as texturing of the material \cite{kalidindi1992} or damage nucleation \cite{Kovacik1999} are not incorporated into the simulation. The simulation is time dependent, with non-linear geometry so that finite strains and rotations are accounted for. We do not introduce any geometrical imperfections to trigger necking, but instead rely upon dynamic waves propagating through the material to trigger the instability. Note that we are intending to approximate a quasi-static deformation, but these dynamic waves (of small magnitude) are simply a convenient way to trigger the instability, and the inertial forces arising from the acceleration terms are otherwise negligible. We use CPS4R elements which correspond to two-dimensional square (in the undeformed frame), plane stress elements with nodes at each corner. We have $40$ elements in the $X$ direction and $240$ elements in the $Y$ direction. We specify the simulation to run for 1 s with automatic time incrementation. However, at approximately 0.6\,s, which we refer to as $t_f$, the deformation becomes highly localized in individual elements and the simulation is no longer valid. Hence, we do not report results after this point. This approximately corresponds to the point in time when the localized neck would form. The strain output from this forward problem is shown in \cref{fig:strainoutput} for two different points in time. This is used as input for the inverse problem. We output the displacement of each node and the stress in each element at 2000 evenly spaced time intervals. 

\subsection{Inverse problem}
\label{ssec:inverseproblem}
Here, we consider a subset of the domain considered for the forward problem. This allows us to avoid complications arising from the singularities at the corners of the sample, for which the numerical method does not apply. The modified domain is shown in \cref{fig:boundaryconditions} where the length in the $X$ direction is $l'_X$ (1/8 of the domain is removed from each side). We redefine $X$ such that $X=0$ on the left hand side boundary. We have the traction on the $X=0$ and $X=l'_X$ boundaries, obtained directly from the forward problem. When a subset of the characteristics are leaving the domain, we choose the component of the traction to specify such that there is no reflection (see the discussion on boundary conditions in \cite{Cameron2022a}). We choose the direction of information propagation to be  rightward for the approximately horizontal characteristics, and upward for the approximately vertical characteristics (\cref{fig:characteristics}). The reader should refer to \cite{Cameron2021,Cameron2022a} for a discussion of the characteristics. For this particular deformation, the inertial terms in the force balance equation will be negligible, hence, we do not include them in the inverse problem computation. We also use the elastic constants specified in the forward problem (for a real experiment they would have to be determined using one of the approaches discussed in \cref{sec:rangeofapplicability}).  

\begin{figure}[!htb]
\centering
\includegraphics[width=1\textwidth]{{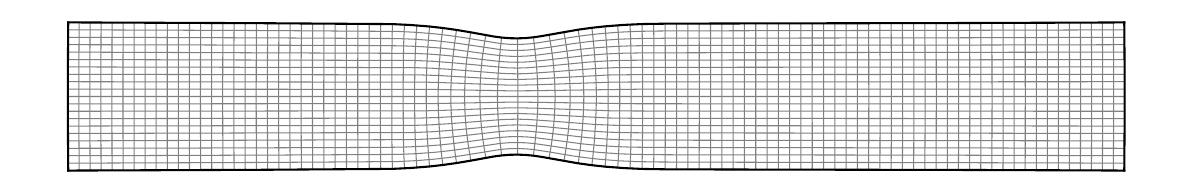}}
\caption{The domain and computational grid used for the inverse problem at time $t=t_f$. Each cell in the figure is four finite volumes in the computation. }
\label{fig:geomdeformed}
\end{figure}

The thickness $h$ is defined as a function of position and time:
\begin{equation}
   h = \bar{h}(\bx,t).
   \label{eq:thicknesscorrection}
\end{equation}
As the elastic volume change will be negligible for this problem, we assume $\mathrm{det}(F_{ij})=1$. This gives $h/h_0 = 1/\mathrm{det}(F_{\alpha \beta})$ at every point in the sample throughout the deformation, where $h_0$ is the initial thickness. Note we use Greek letters ($\alpha,\beta,...$) to indicate the index when it takes values in the set $\{1,2\}$, and Latin letters ($i,j,...$) to indicate the index when it takes values in the set $\{1,2,3\}$. Therefore, $\mathrm{det}(F_{\alpha \beta})$ corresponds to the determinant of the two by two matrix in the $(x,y)$ plane (the (3,3) position in the cofactor matrix). We use a modified form of of the governing equations to account for varying thickness using the variable $\bT'$ where $\bT' = h\bT$:
\begin{subequations}
\begin{gather}
    T_{\alpha\beta}' = h T_{\alpha\beta} \label{eq:temp30}\\
    \frac{\partial T'_{\alpha\beta}}{\partial x_\beta} = 0,\\
    \dot{q}^{(\alpha)}_\gamma =  \frac{q^{(\beta)}_\gamma q^{(\beta)}_\delta D_{\delta \zeta}q^{(\alpha)}_\zeta(\lambda^{e(\beta)}_{\bB}+\lambda^{e(\alpha)}_{\bB})}{\lambda^{e(\beta)}_{\bB}-\lambda^{e(\alpha)}_{\bB}}
        +W_{\gamma\delta}q^{(\alpha)}_\delta\quad \mathrm{where}\quad \alpha \neq \beta,
        \label{eq:forceballance2Dc}\\
     q^{(\alpha)}_3 = 0,\\
    T_{\alpha \beta}' = \sum_\gamma\sigma^{\prime (\gamma)}q_\alpha^{(\gamma)} q_\beta^{(\gamma)}, \qquad B_{ij}^e = \sum_k\lambda^{e(k)}_{\bB}q_i^{(k)}q_j^{(k)},\\
    B_{ij}^e = f_{ij}(\{T_{kl}\}).
\end{gather}
\label{eq:forceballance2D}
\end{subequations}

Here we implement an algorithm that is first order accurate in time. We use $j$ to specify the increment in time and $n_t$ for the total number of increments in time. We use the subscript to denote the variable evaluated at that increment in time, e.g. $\bT'_{j}$ corresponds to the stress at the $j$ increment.

We do not compute $\dot{\bq}$ directly using \cref{eq:forceballance2Dc}, instead we construct a matrix $\bA$, based on expression \cref{eq:temp13}, which will have the same eigenvectors as the stress up to first order accuracy:
\begin{equation}
    \bA_{j} = \bF^*_j\bB^e_{j-1}\bF^{*\top}_j,
    \label{eq:temp23}
\end{equation}
where $\bF^*_j=\bF_{j}\bF_{j-1}^{-1}$. We can then directly compute the eigenvectors of $\bA$ and hence the eigenvectors of $\bT$. This is done for each material point.

$F_{\alpha\beta}$ is computed for each element at each time step using the displacements from the nodes at the corners of the element. Specifically, linear regression is used to fit the displacement gradient as all 4 nodes overdetermine it. $\bF^*$, $\bB$, and $h$ are computed from $F_{\alpha\beta}$. Once the eigenvectors $\{\bq_{j}^{(i)}\}$ are obtained from the computed $\bA_{j}$, the finite volume method developed in \cite{Cameron2021} is used to obtain the stress as a function of position. This stress is then used to compute the elastic strain using the elastic moduli and linear elasticity relationship specified for the forward problem.
After this, we proceed to time step $j+1$. The algorithm used is as follows:

\begin{algorithm}[H]
\SetAlgoLined
%\KwResult{$\bT$ as a function of position and time }
 $j \leftarrow 0$\;
 $\bT_{j} \leftarrow 0$\;
 $j \leftarrow 1$\;
 compute $\{\bq^{(i)}_{j}\}$ by solving $\bD_j\bq^{(i)}_j = \lambda^{(i)} \bq^{(i)}_j$ for all $\mathbf{X}$\;
 \While{$j\leq n_t$}{
 compute $\bT'_{j}$ from $\{\bq_{j}\}$ and boundary conditions for all $\mathbf{X}$ using the finite volume method\;
 compute $\bT_{j}$ from $\bT'_{j}$ using \cref{eq:temp30} for all $\mathbf{X}$\;
 compute $\bF_j$ and $\bF^*_{j+1}$ for all $\mathbf{X}$ using the observed nodal displacements\;
 compute $\bB^e_{j}$ from $\bT_{j}$ for all $\mathbf{X}$ using \cref{eq:linelasticBe}\;
 compute $\bA_{j+1}$ for all $\mathbf{X}$ using \cref{eq:temp23}\;
 compute $\{\bq^{(i)}_{j+1}\}$ by solving $\bA_{j+1}\bq^{(i)}_{j+1} = \lambda^{(i)} \bq^{(i)}_{j+1}$ for all $\mathbf{X}$\;
 $j\leftarrow j+1$\;
 }
 \caption{Compute $\bT$ as a function of position and time}
\end{algorithm}

Kinematic data is output from the forward problem for 2000 time steps, however, as mentioned previously, the deformation becomes highly localized in individual elements and is not valid at times after approximately 1220. Hence, we do not report values at later steps. Furthermore, there is close to zero deformation in the first few time steps, as $d^2\xi(t)/dt^2 = d\xi(t)/dt = 0$ at $t=0$. The algorithm is not robust for this case, hence, we begin the simulation at time step 20, and end at 1220. The number of steps $n_t=1200$ correspond to each set of output data form the forward problem. We also consider the problem with larger time steps ($n_t = 240,120,24$ and $12$) in order to evaluate how the error depends on this value. The algorithm, including the finite volume method, was implemented in Python and utilizes the Scipy sparse linear algebra package \cite{Jones2001}.  The computation with 1200 time steps took approximately 4 hours on a 2019 Macbook Pro with a 2.6 GHz Intel core i7 processor (the computational time was approximately proportional to $n_t$).

\subsection{Results}
\label{ssec:results}
\begin{figure}[!htb]
\centering
\includegraphics[width=1\textwidth]{{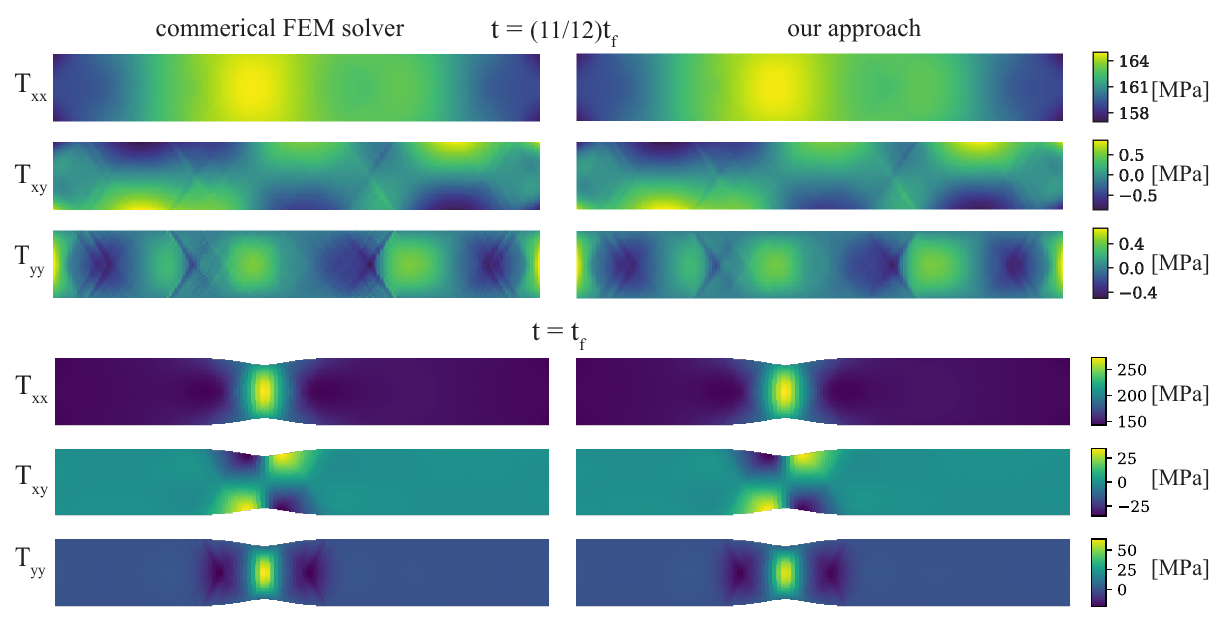}}
\caption{Comparison of the stress field output by the forward problem and inverse problem at time $t = (11/12)t_f$ and $t = t_f$.}
\label{fig:stressfieldcompare}
\end{figure}

\begin{figure}[!htb]
\centering
\includegraphics[width=0.7\textwidth]{{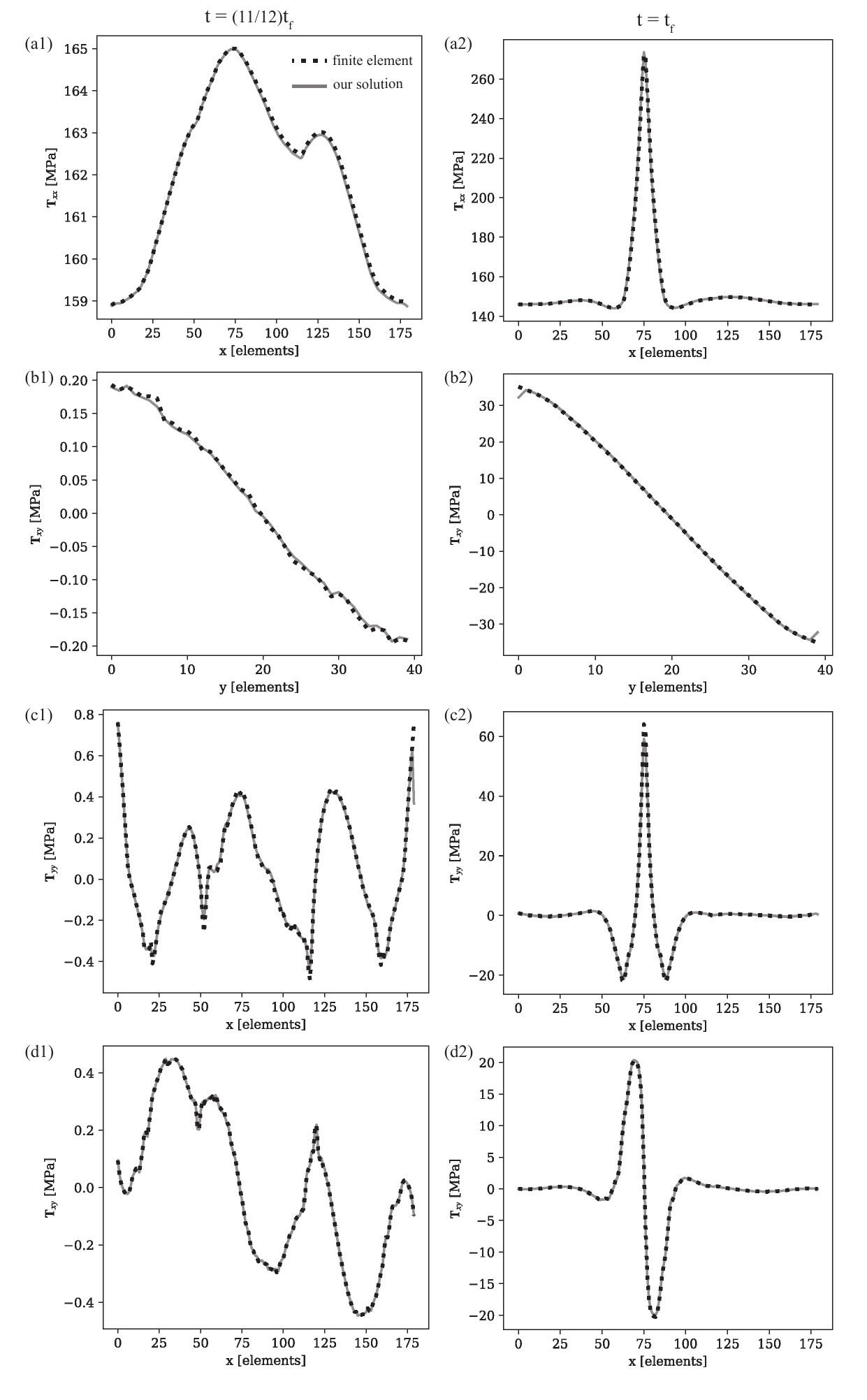}}
\caption{Line comparison of the stress output by the forward problem and inverse problem at time $t = (11/12)t_f$ (a1,b1,c1,d1) and $t = t_f$ (a2,b2,c2,d2). (a) $T_{xx}$ stress component along $Y = l_Y/2$. (b) $T_{xy}$ stress component on a $X=(7/18)l_X$ line that passes through the neck. This was chosen to be close to the center, but not at the center so the shear stress was non-zero. (c) $T_{yy}$ component along $Y = l_Y/2$, (d) $T_{xy}$ stress component on a $Y=l_Y/4$ line. }
\label{fig:stresslinecompare}
\end{figure}

We compare the stress field computed from the inverse problem with the stress field from the forward problem in \cref{fig:stressfieldcompare,fig:stresslinecompare}. As can be seen, the stress distributions are in close agreement, indicating close to exact results and demonstrating the validity of the approach.

\begin{figure}[ht!]
\centering
\includegraphics[width=1\textwidth]{{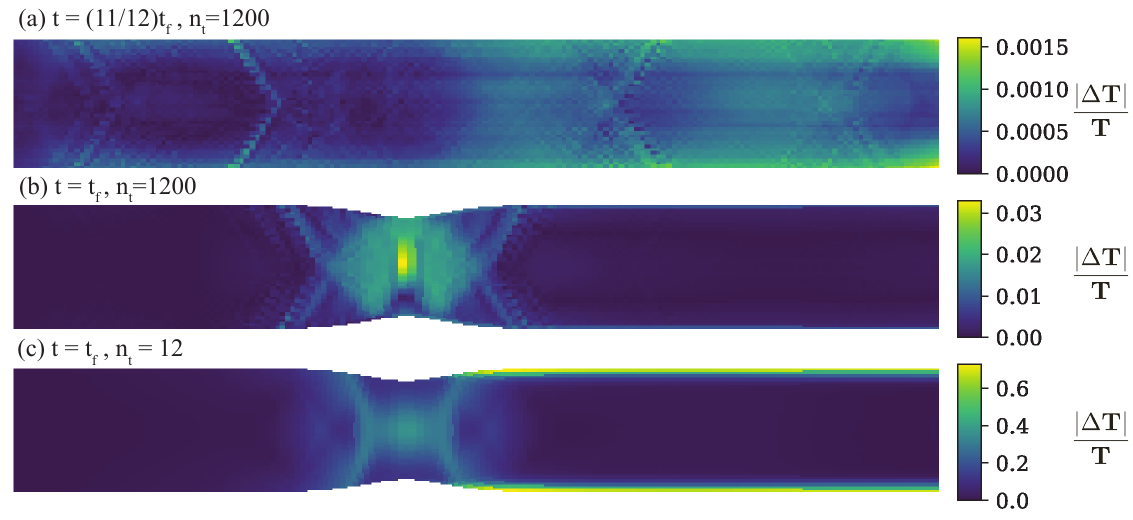}}
\caption{Error as a function of position when compared to the forward problem, plotted at different times for solutions computed using different temporal resolutions.}
\label{fig:errorfield}
\end{figure}

We quantify the error as a function of position using $|\Delta \bT|/\mathrm{mean}_\bx({\bT})$, where $\Delta \bT$ is the difference between the stress computed from the forward problem and the stress computed from the inverse problem, and the $\mathrm{mean}_\bx$ is taken over all positions (but only at the time considered). The $\bT$ used in the denominator is the $\bT$ from the forward problem. We show how the error is spatially distributed in \cref{fig:errorfield}. As can be seen in \cref{fig:errorfield}a, the error increases toward the right hand side. This is because the boundary condition is specified on the left hand side and as the solution is computed along characteristics going from left to right, the numerical error accumulates. The error increases near the boundary to the right of the neck. This is due to the characteristic curves not being perfectly parallel to the boundary, and is particularly apparent for large time-steps. (The characteristic curves are discussed further later in this section). This error accumulation along the boundaries is discussed in more extensively in \cite{Cameron2022a}. Finally, we note that we see the larger errors in the center of the neck. This is likely because $\{\bq^{(i)}\}$ is changing more rapidly with time, and the finite time step will cause a larger error. 

We quantify the average error as a function of time using the normalized mean absolute error (NMAE):
\begin{equation}
    \mathrm{NMAE}(\bT) = \frac{\mathrm{mean}_\bx(|\Delta \bT|)}{\mathrm{mean}_\bx(|\bT|)}.
\end{equation}
Note that elements directly adjacent to the boundary are not included in the error computation. There is moderate error accumulation at the boundary elements when using the current numerical approach, but this can be removed during post processing when using the data to investigate real materials or deformations. Hence, the authors considered it more representative not to include this in the overall error metric (if included the boundary error would largely control the metric). The reader can observe the magnitudes of these errors in \cref{fig:stresslinecompare}\, b2,c1, and can also refer to \cite{Cameron2022a} for a more extensive discussion on this issue.

\begin{figure}[!htb]
\centering
\includegraphics[width=1\textwidth]{{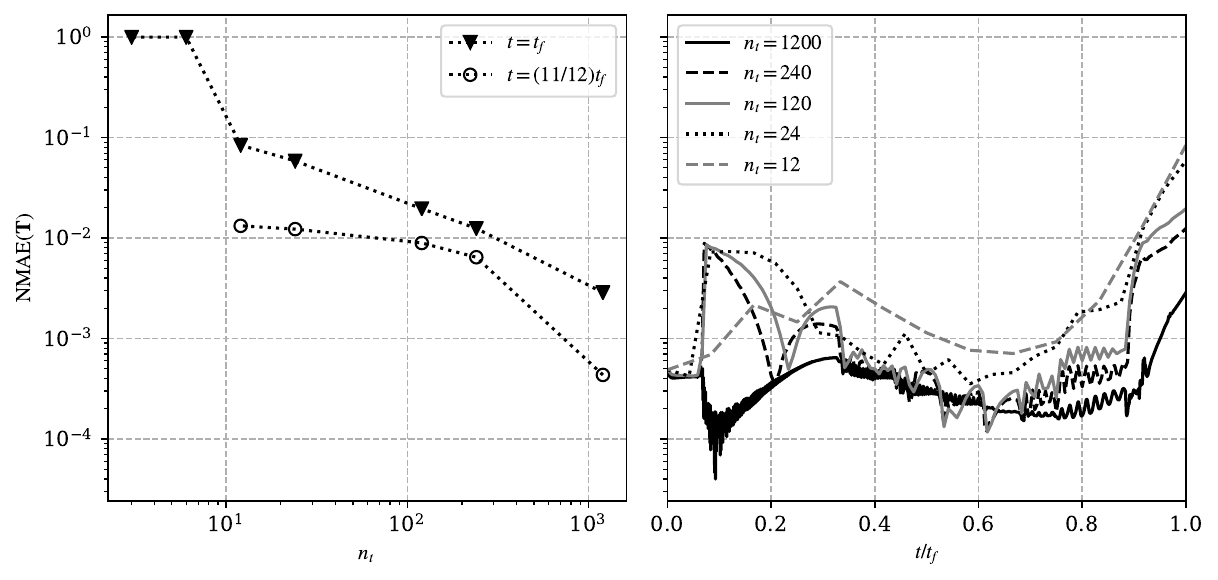}}
\caption{NMAE. (a) Error as a function of the resolution at two different times. It can be seen that increasing the number of steps $n_t$ is likely to improve accuracy further. (b) Error as a function of time for five different values of $n_t$. It can be seen that error increases substantially near the end where necking and significant rotations start to occur.  }
\label{fig:error}
\end{figure}

The NMAE was 2.1$\times 10^{-3}$ at the time when the simulation was terminated (this was the point in time with the maximum NMAE). The error is plotted as a function of time for different sizes of time-step (\cref{fig:error}). As can be seen in \cref{fig:error}a, the error decreases as the number of temporal increments are increased. There are several possible sources for error: the temporal resolution, the spatial resolution, the approximation of constant volume when computing the thickness, the inertial terms in the force balance equation, and error in the forward problem computation. As the error is decreasing with increasing $n_t$, this indicates that the temporal resolution is the primary source of error. This is particularly true at higher values of $t$ where necking occurs. It is likely that with further reduction of $n_t$, the error would decrease further. 

\begin{figure}[!htb]
\centering
\includegraphics[width=0.75\textwidth]{{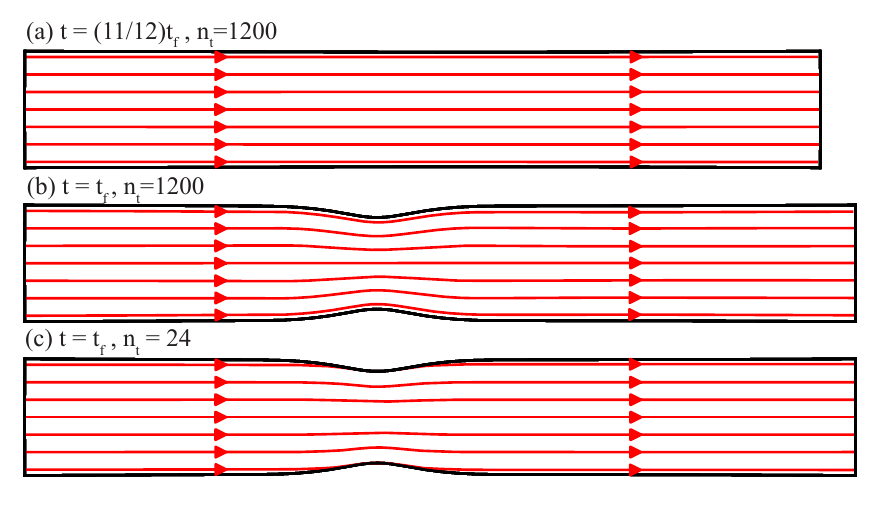}}
\caption{Spatial characteristic curves obtained when solving the inverse problem. These are obtained using an ODE solver and the eigenvectors. There is an additional set of characteristic curves at right angles with information propagating approximately in the upward direction.  }
\label{fig:characteristics}
\end{figure}

The characteristic curves, curves where the tangent is aligned with the eigenvectors of the stress at each point in space, play a critical role in determining how the solution is computed (see \cref{ssec:simplifiedequations}). Information is propagated along these curves as the solution is computed at each time step. For visualization purposes, the characteristic curves corresponding to the stress computation at specific times are computed using an ODE solver (\cref{fig:characteristics}). As the computation is two-dimensional, there are two sets of characteristic curves corresponding to each principal direction. The horizontal characteristic curves are shown, and the arrows indicate the arbitrarily chosen direction of information propagation used in the computational procedure. As can be seen from \cref{fig:characteristics}c, when the time step is large, there is substantial error in the characteristic curves. The curves should be aligned with the the boundary at the boundary, as there will be no shear traction applied. This error diminishes as the time step is made smaller. 

\begin{figure}[!htb]
\centering
\includegraphics[width=0.8\textwidth]{{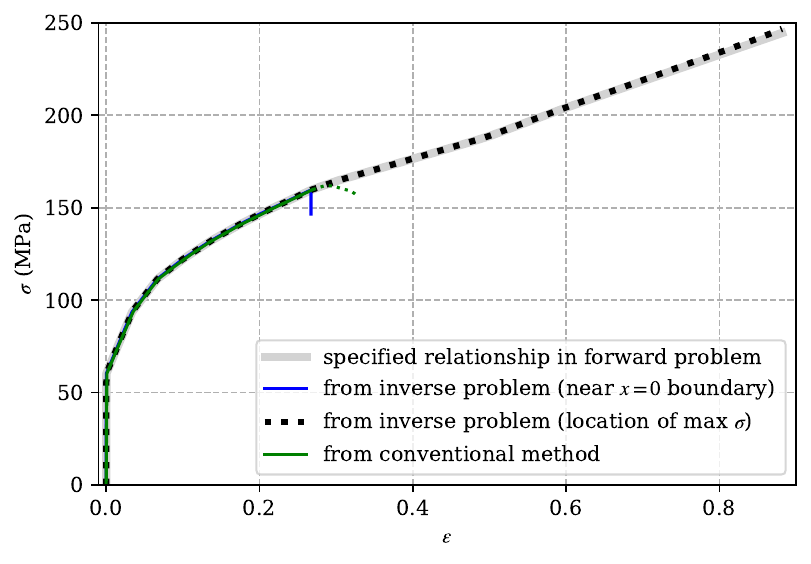}}
\caption{Stress-strain curve for material elements obtained via the inverse problem, compared to the specified relationship in the forward problem. The von Mesis stress $\sigma = \sqrt{3/2}|\bT_0|$ and the accumulated plastic strain $\varepsilon$. The green stress-strain curve shows what would be obtained using a conventional analysis that assumes a uniform cross section (the dotted line corresponds to when the assumption breaks down). Note the lines directly overlay each other until necking and some are not visible. }
\label{fig:stressstrain}
\end{figure}

The approach developed in this article could be used to obtain the stress-strain relationship for a material at high strains (i.e. by plotting the stress and strain for an element in the neck). We show this plot for the material element subject to the highest stress in \cref{fig:stressstrain}. This is compared with the relationship specified in the forward problem, and one can see they are a close match. When plotting the stress-strain relationship for a material element near the $X=0$ boundary, we see the elastic unloading that occurs at the same time the neck forms in the center of the sample. We contrast these curves with the stress strain relationship that would be obtained using the conventional method used to analyze tensile specimens: measuring $\Delta l$ along the gauge, calculating the engineering strain, computing the engineering stress from the force, and calculating the true stress and strain by assuming a constant cross-section. The methods are both accurate for strain values up to necking ($\varepsilon=0.26$). However, information at higher strains is not available using the approach which assumes a constant cross section. The approach presented in this paper gives accurate stress-strain behavior up to $\varepsilon=0.90$, this corresponds to an increase in length of a material element of 142\%. Other approaches may be applied to estimate the stress in the neck, as discussed earlier in this section, however, more restrictive constitutive equation assumptions must be made, and significant errors still arise as the solutions are inexact. For example, the Bridgman correction (applicable to cylindrical geometries), gives errors on the order of 10\% for strains on the order of $70-80\%$.  Of course, before application to real materials, the sensitivity to erroneous kinematic fields must be investigated, along with the validity of additional assumptions such as constant elastic modulus, and the isotropy of the material.

\section{Conclusion}
The system of partial differential equations previously developed \cite{Cameron2021,Liu2021} enabled one to deterministically obtain the stress in the cases of isotropic elastic or pure plastic deformation. However, it was not possible to apply this to determine material properties in many simple cases, such as a necking tensile specimen, because some regions elastically deform while others undergo plastic deformation. The system of equations derived in this article (\cref{eq:systemall}) broadens the applicability of this approach to include these more complex isotropic elasto-plastic deformations, addressing a significantly wider range of cases. Although, as discussed, there will be some cases where the system of equations give non unique solutions. Numerical solutions to the equations are demonstrated using strain fields obtained from finite element simulations, and the results obtained are validated against the same simulations (the NMAE at the final time step of the simulation was $2.1\times10^{-3}$).

More research is needed before this approach can become a widely used tool. This includes developing equations applicable to anisotropic material models, robustly addressing cases of non-unique stress eigenvectors, and addressing practical issues such as noisy experimental datasets.

\bibliographystyle{unsrt} 
\bibliography{references}
\appendix
\appendixpage
\section{Large deformation coaxiality}
\label{sec:finitecoaxiality}
% [no repeated eigen values][Dp Neq 0]
% language coaxial, communte, eigenvectors, elastoplastic
% Given the Kröner-Lee multipicative decomposition $\bF = \bF^e\bF^p$ we introduce the reference body, the structural space, and the deformed body.

We show that  $\bT$ is coaxial with $\bF^e\bD^p\bF^{e\top}$ provided the second elastic Piola-Kirchhoff stress $\bT^e$ is coaxial with the plastic flow $\bD^p$, and the right Cauchy stress tensor $\bC^e$, as discussed in \cref{ssec:coaxiality} (these variables are defined in the structural reference configuration). As discussed in \cref{ssec:coaxiality}, $\bD^p$ will be coaxial with $\bT^e$ in the case where both have unique eigenvectors and the function mapping between them is isotropic and invariant under reflection. However, there may be cases where $\bD^p$ has repeated eigenvalues and non-unique eigenvectors. A common example is when the sample has not yielded and $\bD^p = 0$. Nevertheless, one can choose from the set of non-unique eigenvectors, the eigenvectors that are the same as those of $\bT^e$ and $\bC^e$. This will always be possible provided the function mapping between them is isotropic and invariant under reflection.

It is convenient to utilize the singular value decomposition of $\bF^e$:
\begin{equation}
    \bF^e = \sum_i\lambda_\bF^{e(i)}\bl^{e(i)}\br^{e(i)\top} = \bZ_L \bLambda_\bF^e \bZ_R^{\top},
\end{equation}
where $\{\lambda_\bF^{e(i)}\}$ are the elastic principal stretches (singular values) of $\bF^e$, $\{\bl^{e(i)}\}$ are the left principal directions, $\{\br^{e(i)}\}$ are the right principal directions, $\bLambda_\bF^e$ is a diagonal matrix which contains the elastic principal stretches, $\bZ_L$ is an orthonormal matrix with columns $\{\bl^{e(i)}\}$ and $\bZ_R$ is an orthonormal matrix with columns $\{\br^{e(i)}\}$.

Given its definition, $\bC^e$ will have eigenvectors $\{\br^{e(i)}\}$ and hence can be expressed as $\bC^e = \bZ_R\bLambda_\bC^e\bZ_R^\top$. Due to coaxiality, $\bT^e$ and $\bD^p$ will also have eigenvectors $\{\br^{e(i)}\}$ and can be diagnonalized using $\bZ_R$, i.e. $\bT^e = \bZ_R\bLambda_\bT^e\bZ_R^\top$ and $\bD^p = \bZ_R\bLambda_\bD^p\bZ_R^\top$. 

Inverting the definition of $\bT^e$ (\cref{eq:temp35}) we have:
\begin{equation}
    \bT =\frac{1}{\mathrm{det}\bF^e}\bF^e\bT^e\bF^{e\top}.
\end{equation}
Substituting the diagonalized forms of these quantities we have:
\begin{equation}
    \bT =\frac{1}{\mathrm{det}\bF^e}
    (\bZ_L\bLambda_\bF^e \bZ_R^{\top})(\bZ_R\bLambda_\bT^e\bZ_R^\top)(\bZ_R \bLambda_\bF^e \bZ_L^{\top}).
\end{equation}
Utilizing the fact that $\bZ_R^{\top}\bZ_R = \bI$ gives:
\begin{equation}
    \bT = \frac{1}{\mathrm{det}\bF^e}
    (\bZ_L\bLambda_\bF^e \bLambda_\bT^e\bLambda_\bF^e \bZ_L^{\top}).
    \label{eq:temp26}
\end{equation}
Hence, $\bT = \bZ_L \bLambda_\bT\bZ_L^\top$ and will have eigenvectors $\{\bl^{e(i)}\}$. 

Using the same argument, we can show that the quantity $\bF^e\bD^p\bF^{e\top}$ has $\{\bl^{e(i)}\}$ as eigenvectors:
\begin{subequations}
\begin{gather}
   \bF^e\bD^p\bF^{e\top} =  (\bZ_L\bLambda_\bF^e \bZ_R^{\top})(\bZ_R\bLambda_\bD^p\bZ_R^\top)(\bZ_R \bLambda_\bF^e \bZ_L^{\top}),\\
   \bF^e\bD^p\bF^{e\top} = \bZ_L\bLambda_\bF^e\bLambda_\bD^p\bLambda_\bF^e\bZ_L^{\top}.
\end{gather}
\label{eq:temp27}
\end{subequations}

Finally, we can trivially show that $\bB^e = \bF^e\bF^{e\top}$ has the same eigenvectors:
\begin{subequations}
\begin{gather}
    \bB^e = (\bZ_L\bLambda_\bF^e \bZ_R^{\top})(\bZ_R \bLambda_\bF^e \bZ_L^{\top}),\\
    \bB^e = \bZ_L(\bLambda_\bF^e)^2\bZ_L^{\top}.
\end{gather}
\label{eq:temp28}
\end{subequations}

Together \cref{eq:temp26,eq:temp27,eq:temp28} show that $\bT$, $\bB^e$ and $\bF^e\bD^p\bF^{e\top}$ are all coaxial when the stated assumptions are valid.

Finally, we note that it can be shown that $\bF^{e-\top}$ has the same left and right principal directions as $\bF^e$. Similarly, $\bF^{e-1}$ has the same left and right principal directions as $\bF^{e\top}$. Hence, the above argument could be repeated swapping these variables, showing that the specific combination $\bF^e\bD^p\bF^{e\top}$ is not unique. 

\section{Infinitesimal deformation governing equations}
\label{sec:infintesimalderivation}
Here we give a derivation of the governing elasto-plastic deformation equations in the simplified case of infinitesimal strain. This argument follows parallel to that presented in \cref{ssec:governingequations}. We also note that the same result can be obtained by simplifying the more general finite deformation equations (see \cref{ssec:limitingcases}).

The deformation is considered from time $t=\tau$ to time $t = \tau+\Delta t$. It is assumed that the Cauchy stress $\bT$ is known at time $\tau$ and we wish to determine the stress at time $\tau+\Delta t$. Subscripts are introduced to denote quantities evaluated at these times, e.g. $\bT_{\tau} = \bar{\bT}(\tau)$ and $\bT_{\tau+\Delta t} = \bar{\bT}(\tau+\Delta t)$. We define $\bE^*$ to correspond to the deformation from $\tau$ to $\tDt$, i.e we have:
\begin{equation}
    \bE_\tDt = \bE_\tau+\bE^*.
\end{equation}
We define $\bE^{p*}$ in a similar way:
\begin{equation}
    \bE^p_\tDt = \bE^p_\tau+\bE^{p*}.
\end{equation}
Substituting these expressions into the additive decomposition for strain (\cref{eq:temp18}) gives:
\begin{equation}
    \bE^*+\bE^e_\tau+\bE^p_\tau=\bE^e_\tDt + \bE^{p*} + \bE^p_\tau,
\end{equation}
\begin{equation}
    \bE^*+\bE^e_\tau=\bE^e_\tDt + \bE^{p*}_\tau.
    \label{eq:temp19}
\end{equation}
We express the spectral decomposition of $\bT$ as:
\begin{equation} 
\bT = \sum_i \sigma^{(i)}\bq^{(i)}\bq^{(i)\top},
\end{equation}
where $\{\bq^{(i)}\}$ are the orthonormal eigenvectors of \bT, $\{\sigma^{(i)}\}$ are the eigenvalues, and the superscript  $i\in\{1,2,3\}$. We also assume the eigenvalues do not repeat. Note that $\bE^e$, $\bEd^p$ and $\bT$ have the same $\{\bq^{(i)}\}$ (see \cref{ssec:preliminaries}). Hence, we have:
\begin{equation}
\label{eq:temp3}
    \bE^e\bq^{(i)} =  \lambda^{(i)}\bq^{(i)},
\end{equation}
\begin{equation}
\label{eq:temp4}
    \dot{\bE}^{p}\bq^{(i)} = \gamma^{(i)}\bq^{(i)}.
\end{equation}
where $\{\bq^{(i)}\}$ are the same eigenvectors and $\{\gamma^{(i)}\}$ are the eigenvalues. We consider the expression:
\begin{equation}
    \bE^e_\tDt+\dot{\bE}^p_\tDt\Dt.
\end{equation}
This will have the same eigenvectors as $\bT_\tDt$ as the sum of two tensors with the same eigenvectors also will have the same eigenvectors. The eigenvectors are also invariant to multiplication of the tensor by a scalar quantity. Hence we have:
\begin{equation}
    (\bE^e_\tDt+\dot{\bE}^p_\tDt\Delta t)\bq_\tDt^{(i)} =  (\lambda_\tDt^{(i)}+\Dt\gamma_\tDt^{(i)})\bq_\tDt^{(i)}.
    \label{eq:temp37}
\end{equation}
In addition, we note that:
\begin{equation}
    \bE^{*} = \dot{\bE}_\tDt\Delta t + o(\Dt),
    \label{eq:temp22}
\end{equation}
\begin{equation}
    \bE^{p*} = \dot{\bE}^p_\tDt\Delta t + o(\Dt).
\end{equation}
Hence, we have:
\begin{equation}
    (\bE^e_\tDt+\bE^{p*} + o(\Dt))\bq_\tDt^{(i)} +  =  (\lambda_\tDt^{(i)}+\Dt\gamma_\tDt^{(i)})\bq_\tDt^{(i)}.
\end{equation}
Combining this with \cref{eq:temp19} then gives:
\begin{equation}
    (\bE^*+\bE^e_\tau+ o(\Dt))\bq_\tDt^{(i)}=  (\lambda_\tDt^{(i)}+\Dt\gamma_\tDt^{(i)})\bq_\tDt^{(i)}.
    \label{eq:temp21}
\end{equation}
In the case where $\Dt$ is small, this gives an expression for the eigenvectors of $\bT_\tDt$ in terms of the observed quantity $\bE^*$ and $\bE^e_\tau$. The latter of which can be computed from $\bT_\tau$ using the assumption that the elastic component of the constitutive equation is known (\cref{ssec:assumption}). We can use this expression to derive a differential equation for the evolution of $\{\bq^{(i)}\}$. 

Substituting \cref{eq:temp22} into \cref{eq:temp21} gives:
\begin{equation}
    \bE^e_\tau\bq_\tDt^{(i)}+\Dt\dot{\bE}_\tDt\bq_\tDt^{(i)}+o(\Dt) = (\lambda_\tDt^{(i)}+\Dt\gamma_\tDt^{(i)})\bq_\tDt^{(i)}
\end{equation}
From this, we subtract \cref{eq:temp37} evaluated at time $t=\tau$, giving:
\begin{equation}
    \bE^e_\tau(\bq_\tDt^{(i)}-\bq_\tau^{(i)})+\Dt\dot{\bE}_\tDt\bq_\tDt^{(i)}+o(\Dt) =
    (\lambda_\tDt^{(i)}\bq_\tDt^{(i)} - \lambda_\tau^{(i)}\bq_\tau^{(i)})
    +\Dt\gamma_\tDt^{(i)}\bq_\tDt^{(i)}
\end{equation}
Dividing through by $\Dt$ and taking the limit as $\Dt
\rightarrow 0$ gives:
\begin{equation}
    \bE^e\dot{\bq}^{(i)}+\dot{\bE}\bq^{(i)} =
    \frac{d}{dt}(\lambda^{(i)}\bq^{(i)})
    +\gamma^{(i)}\bq^{(i)}
\end{equation}
\begin{equation}
    \bE^e\dot{\bq}^{(i)}+\dot{\bE}\bq^{(i)} = \lambda^{(i)}\dot{\bq}^{(i)} +
    (\dot{\lambda}^{(i)}+\gamma^{(i)})\bq^{(i)}
\end{equation}

We now multiply on the left hand side by $\bq^{(j)\top}$ for $j\neq i$, noting that $\bq^{(j)\top}\bq^{(i)}=0$ as $\bE^e$ is symmetric and the eigenvectors are orthogonal:
\begin{equation}
    \bq^{(j)\top}\bE^e\dot{\bq}^{(i)}+\bq^{(j)\top}\dot{\bE}\bq^{(i)} = \lambda^{(i)}\bq^{(j)\top}\dot{\bq}^{(i)}.
\end{equation}
\begin{equation}
    \bq^{(j)\top}(\bE^e-\lambda^{(i)}\bI)\dot{\bq}^{(i)}+\bq^{(j)\top}\dot{\bE}\bq^{(i)} = 0.
\end{equation}

Substituting in $\bq^{(j)\top}\bE^e = \lambda^{(j)}\bq^{(j)\top}$ and rearranging gives:
\begin{equation}
    \bq^{(j)\top}\dot{\bq}^{(i)} = 
    \frac{\bq^{(j)\top}\dot{\bE}\bq^{(i)}}{\lambda^{(i)}-\lambda^{(j)}},
\end{equation}
for all $j$ such that $j\neq i$.
Recall that we have made the assumption that the eigenvectors of $\bB^e$ are unique, hence $\lambda^{(i)}\neq\lambda^{(j)}$ for $i \neq j$. Since, $\bq^{(j)\top}\bq^{(j)}=1$
 and $\bq^{(j)\top}\bq^{(i)}=0$ for all $j$ such that $j\neq i$, we have:
\begin{equation}
    \dot{\bq}^{(i)} = \sum_{j,j\neq i}
    \frac{\bq^{(j)}\bq^{(j)\top}\dot{\bE}\bq^{(i)}}{\lambda^{(i)}-\lambda^{(j)}}.
\end{equation}

This, combined with force balance and the assumption that the elastic component of the constitutive equation is known gives a full deterministic system of PDEs:
\begin{subequations}
\begin{gather}
    \mathrm{div}{\mathbf{T}} + \mathbf{b} = \rho\ddot{ \mathbf{x}},\\ 
    \ \dot{\bq}^{(i)} = \sum_{j,j\neq i} \frac{\mathbf{q}^{(j)}\mathbf{q}^{(j)\top}\dot{\bE}\mathbf{q}^{(i)}}{\lambda^{e(i)}_\bE - \lambda^{e(j)}_\bE},\\
    \bT = \sum_i\sigma^{(i)}\bq^{(i)}\bq^{(i)\top}, \qquad \bE^e = \sum_i\lambda^{e(i)}_{\bE}\bq^{(i)}\bq^{(i)\top},\\
    \bE^e = \Bar{\bE^e}(\bT) \qquad \mathrm{in} \qquad \Omega,\\
    \bd^{(k)\top}\bT\bn = \bd^{(k)\top}\bt
    \qquad \mathrm{on} \qquad \partial\Omega\\
    \bq^{(i)} = \bq_{t0}^{(i)} \qquad \mathrm{at} \qquad t=t_0.
\end{gather}
\end{subequations}
Similarly to the finite deformation case, these equations are non-linear and time dependent.
\end{document}